\documentclass{SciPost}
\usepackage{graphicx,wrapfig}
\usepackage{comment}
\usepackage{bm}

\usepackage{color}
\def\red{\color{red}}

\def\blue{\color{blue}}

\usepackage[numbers,sort&compress]{natbib}

\usepackage{amsmath,amssymb,mathrsfs}
\usepackage{graphicx}

\DeclareMathOperator{\sign}{sign}

\DeclareMathOperator{\re}{Re}

\newcommand{\la}{\label}
\newcommand{\bbm}{\begin{multline}}
\newcommand{\eem}{\end{multline}}
\newcommand{\be}{\begin{equation}}
\newcommand{\ee}{\end{equation}}
\newcommand{\bea}{\begin{eqnarray}}
\newcommand{\eea}{\end{eqnarray}}
\newcommand{\p}{\partial}

\newcommand{\R} {\mbox{Re}\,}
\newcommand{\I} {\mbox{Im}\,}

\usepackage{color}
\def\red{\color{red}}

\def\blue{\color{blue}}

\begin{document}

% TODO: write your article's title here. 
% The article title is centered, Large boldface, and should fit in two lines
\begin{center}{\Large \textbf{
Odd surface waves in two-dimensional incompressible fluids
}}\end{center}

% TODO: write the author list here. Use initials + surname format.
% Separate subsequent authors by a comma, omit comma at the end of the list.
% Mark the corresponding author with a superscript *. 
\begin{center}
Alexander G. Abanov\textsuperscript{1,2}, 
Tankut Can\textsuperscript{4}, 
Sriram Ganeshan\textsuperscript{3,1}
%Gustavo Monteiro\textsuperscript{3}
\end{center}

% TODO: write all affiliations here. 
% Format: institute, city, country
\begin{center}
{\bf 1} Simons Center for Geometry and Physics, Stony Brook University, Stony Brook, NY 11794, USA \\
{\bf 2} Department of Physics and Astronomy, Stony Brook University, Stony Brook, NY 11794, USA\\
{\bf 3} Department of Physics, City College, City University of New York, New York, NY 10031, USA \\
{\bf 4} Initiative for the Theoretical Sciences, The Graduate Center, CUNY, 10012, USA

% TODO: provide email address of corresponding author
%* CorrespondingAuthor@email.address
\end{center}

\begin{center}
\today
\end{center}

% For convenience during refereeing: line numbers
%\linenumbers

\section*{Abstract}
{\bf 
% TODO: write your abstract here.
We consider free surface dynamics of a two-dimensional incompressible fluid with odd viscosity. The odd viscosity is a peculiar part of the viscosity tensor which does not result in dissipation and is allowed when parity symmetry is broken. For the case of incompressible fluids, the odd viscosity manifests itself through the free surface (no stress) boundary conditions. We first find the free surface wave solutions of hydrodynamics in the linear approximation and study the dispersion of such waves. As expected, the surface waves are chiral and even exist in the absence of gravity and vanishing shear viscosity.  In this limit, we derive effective nonlinear Hamiltonian equations for the surface dynamics, generalizing the linear solutions to the weakly nonlinear case. Within the small surface angle approximation, the equation of motion leads to a new class of non-linear chiral dynamics governed by what we dub the {\it chiral} Burgers equation. The chiral  Burgers  equation is  identical to the complex Burgers equation  with imaginary viscosity and an additional analyticity requirement that enforces chirality. We present several exact solutions of the chiral Burgers equation. For generic multiple pole initial conditions, the system evolves to the formation of singularities in a finite time similar to the case of an ideal fluid without odd viscosity. We also obtain a periodic solution to the chiral Burgers corresponding to the non-linear generalization of small amplitude linear waves.

}

% TODO: include a table of contents (optional)
% Guideline: if your paper is longer that 6 pages, include a TOC
% To remove the TOC, simply cut the following block
\vspace{10pt}
\noindent\rule{\textwidth}{1pt}
\tableofcontents\thispagestyle{fancy}
\noindent\rule{\textwidth}{1pt}
\vspace{10pt}

\tableofcontents
%%%%%%%%%%%%%%%%%%%%%%%%%%%%%%%%%%%%%%%%%%%%%%%%%%%%%%%%%%%%%%%%%%%%%%%%%%%%%%%%%%%%%%%%%%%%%%%

%%%%%%%%%%%%%%%%%%%%%%%%%%%%
%%%%%%%%%%%%%%%%%%%%%%%%%%%%
\section{Introduction.} 
%%%%%%%%%%%%%%%%%%%%%%%%%%%%

 Recently there has been much interest in the role of parity violating effects in two dimensional incompressible hydrodynamics. This was motivated by the seminal paper by Avron, Seiler, and Zograf \cite{avron1995viscosity} where they showed that the quantum Hall (QH) ground state has non-vanishing non-dissipative odd viscosity. In two dimensions, the odd part of the viscosity tensor is compatible with isotropy and has an elegant interpretation as the adiabatic curvature on the space of flat background metrics~\cite{avron1995viscosity}.  The role of odd viscosity in the context of QH fluids (where it is dubbed Hall viscosity) has been an active area of research~\cite{tokatly2006magnetoelasticity,tokatly2007new,tokatly2009erratum, read2009non,haldane2011geometrical,haldane2011self,hoyos2012hall, bradlyn2012kubo, yang2012band,abanov2013effective,hughes2013torsional, hoyos2014hall, laskin2015collective, can2014fractional,can2015geometry,klevtsov2015geometric,klevtsov2015quantum, gromov2014density, gromov2015framing, gromov2016boundary, scaffidi2017hydrodynamic, andrey2017transport,alekseev2016negative,pellegrino2017nonlocal}.
  
 Avron subsequently considered the case of classical 2D hydrodynamics with dominant odd viscosity~\cite{avron1998odd}. Recent works have further outlined observable consequences of the odd viscosity in classical two dimensional incompressible hydrodynamics ~\cite{wiegmann2014anomalous,lapa2014swimming, banerjee2017odd, ganeshan2017odd,lucas2014phenomenology}. In three dimensions, the general parity odd terms of the viscosity tensor were considered in the context of a plasma in a magnetic field~\cite{landau1987fluid} and in hydrodynamic theories of superfluid He-3A \cite{helium-book}.

 In this work, we consider the classical problem of deep water surface waves with the addition of odd viscosity. For simplicity, we refer to these as simply {\it odd surface waves}.  The generalization of this work to the odd version of shallow water surface waves is straightforward. Our starting point is the incompressible 2D Navier-Stokes equation with free boundary and in the presence of odd viscosity~\cite{ganeshan2017odd}. These are basically waves on the 1D surface of a 2D fluid with broken parity symmetry. We parameterize the boundary of the fluid using a height function defined by $y=h(x,t)$ assuming that the fluid domain is defined by $y\leq h(x,t)$. We begin by solving the linearized (neglecting advective acceleration terms) 2D Navier-Stokes equation in the presence of odd viscosity ($\nu_o$) subject to no-stress boundary conditions.  These linearized solutions are the (``odd'') generalization of  linearized viscous Lamb surface wave solutions (in the following just ``Lamb's solutions") with shear viscosity ($\nu_e$) and non-zero vorticity ~\cite{lamb1932hydrodynamics}. The presence of odd viscosity completely changes the surface phenomenon and allows for chiral dispersing waves in the limit of no gravity ($g=0$) and vanishing shear viscosity $\nu_e\rightarrow 0$. The dispersion relation in this limit is of the form 
\be
	\Omega(k)=-2\nu_o k|k|
 \la{eq:OmegaDisp}
\ee 
and is reminiscent of the famous Benjamin-Davis-Ono (BDO) dispersion relation.

Vorticity plays a significant role in the structure of linearized solutions in both Lamb's case and its odd viscosity generalization. In the limit of $\nu_e\rightarrow 0$, the vorticity is confined to a thin layer at the boundary. The thickness of this boundary scales as $\delta\sim \sqrt{\frac{\nu_e}{\nu_o}}$, similar to Lamb's case~\cite{lamb1932hydrodynamics}. However, the presence of odd viscosity significantly alters the scaling of vorticity within the boundary layer, which diverges with vanishing shear viscosity ($\omega \sim \frac{1}{\sqrt{\nu_e}}$) as opposed to a constant vorticity $\omega \sim O(1)$ for Lamb's case. Outside of this layer the vorticity is negligible and the fluid can be approximated by an irrotational fluid which is completely determined by a scalar potential~\cite{longuet1953mass, longuet1963generation}. Ruvinsky et al.~\cite{ruvinsky1991numerical} used this almost irrotational nature of Lamb's solutions with only shear viscosity and gravity and constructed a scheme dubbed the ``quasi-potential approximation'' (QPA) to capture the non-linear surface dynamics. The basic idea of QPA is to integrate out the vortical boundary layer and rewrite it as an effective Bernoulli's equation at the surface in terms of a scalar potential term. The integrated out vortical part modifies the boundary condition which determines the modified pressure at the boundary~\cite{fedorov1998nonlinear, dias2008theory, zakharov2012free}. The boundary layer in incompressible dissipation free flows and the related non-analytic dispersion had been also discussed in the theory of edge modes in fractional Quantum Hall systems~\cite{wiegmann2012nonlinear}.

Following the QPA scheme of Ruvinsky et al., we obtain non-linear potential flow equations for a fluid with odd viscosity. As mentioned earlier, the presence of odd viscosity results in diverging vorticity $\omega \sim \frac{1}{\sqrt{\nu_e}}$ within the boundary layer compared to a constant vorticity in the analysis of Ref. \cite{ruvinsky1991numerical} for shear viscosity. This altered scaling significantly changes the structure of the resulting non-linear equation compared to the one obtained by Ruvinsky et al. for the gravity waves. We derive one-dimensional Hamiltonian equations governing surface dynamics corrected by dispersive terms dependent on odd viscosity. This Hamiltonian dynamics within a small surface angle approximation~\cite{kuznetsov1994formation} takes the form of the complex Burgers equation
\begin{align}
	u_t + 2uu_x - 2i\nu_o u_{xx} = 0,
  \label{eq:almostBO}
\end{align}
where $u(x,t)$ is a complex function. The equation (\ref{eq:almostBO}) is supplemented by an additional condition that $u(z,t)$ is analytic in the lower half plane of $z$. This analyticity condition enforces chiral dynamics at the surface and  selects only special class of solutions from the ones of the simple complex Burgers case. Hence, we dub the complex Burgers with the analyticity condition as {\it chiral Burgers} equation from hereon. 

The above equation can be transformed to Schrödinger equation $i\Psi_t=2\nu_o \Psi_{xx}$ using Cole-Hopf transformation $u=2i\nu_o \Psi_x/\Psi$, where we can identify $\nu_o=-\hbar/4m$. Note that this is chiral Schrödinger equation since $\Psi$ satisfies  the analyticity condition similar to $u$ which makes the dynamics chiral. This is analogous to the transformation that connects Burgers equation to the diffusion equation.

The solution of this equation defines the small angle approximation to the non-linear velocity profile of the fluid.  We analyze the dynamics and present some exact solutions of this effective non-linear equation.  The chiral Burgers equation without the dispersive term (inviscid Burgers) has been previously obtained for surface waves in Ref.~\cite{kuznetsov1994formation}. The odd viscosity adds the dispersive term and changes the character of formation of some of the singularities studied in \cite{kuznetsov1994formation}. 

This paper is organized as follows. We begin by introducing the general hydrodynamic equations in Sec~\ref{sec:statement} and derive the odd viscosity generalization of Lamb's solutions in Sec~\ref{sec:linoddwaves}. We analyze these solutions in the limit $\nu_e\rightarrow 0$ and obtain the scaling of the boundary layer thickness and  velocity and vorticity profile with respect to $\nu_e$ (Sec~\ref{sec:swodd}). Using this scaling for the linearized solutions, we derive the effective equation (chiral Burgers) governing the non-linear surface dynamics of the fluid with odd viscosity in Sec.~\ref{sec:effheur}. In section \ref{sec:hamiltonian}, we derive the non-linear Hamiltonian structure to the second order and deduce mass and momentum conservation laws. We obtain some exact solutions of this new non-linear equation and end the paper with future directions~(Sec~\ref{sec:nln}). Various technical results supporting the main text are presented in appendices. In particular, we also discuss a one parameter family of non-linear equations (Eq.~\ref{eq:family}), which contains the chiral Burgers equation and the  Benjamin-Davis-Ono (BDO) equation as limiting cases.

%%%%%%%%%%%%%%%%%%%%%%%%%%%%
%%%%%%%%%%%%%%%%%%%%%%%%%%%%
\section{Statement of the problem}
%%%%%%%%%%%%%%%%%%%%%%%%%%%%
\label{sec:statement}
%%%%%%%%%%%%%%%%%%%%%%%%%%%%
\subsection{Hydrodynamic equations}

Let us begin from the main hydrodynamic equations with free boundary and in the presence of odd viscosity. In this paper we assume that temperature does not play any major role and the fluid is  incompressible. Moreover, we assume that the density of the fluid is constant and take it to be unity $\rho=1$. Under these assumptions, the equations of hydrodynamics can be written as:
\bea
	\p_i v_i &=& 0 \,, 
 \la{eq:incom} \\	
 	D_t v_i  &=&  \p_j T_{ij}-\p_i(gy)\,.
 \la{eq:euler}
\eea
Here the summation over repeated indices is assumed ($i, j=1,2$) and $D_t\equiv \p_t+v_{i}\p_i$ is a material time derivative. The first equation is the incompressibility condition. The potential $gy$ is an external gravitational potential. To have a closed system of equations one needs to define a constitutive relation expressing the stress tensor $T_{ij}$ in terms of the velocity of the fluid. We write:
\bea
	T_{ij} &=& -p\delta_{ij} +\nu_e(\p_i v_j+\p_j v_i) +\nu_o (\p_i^*v_j+\p_i v_j^*)\,.
 \la{eq:Tij} 
\eea
The first term of the stress tensor (\ref{eq:Tij}) is the standard isotropic pressure term. The other terms come from the viscosity tensor. The second term takes into account the shear viscosity of the fluid with the coefficient $\nu_e$ known as kinematic shear viscosity. The last term in (\ref{eq:Tij}) is less familiar. This is the odd viscosity term. The effects of this term on surface waves are the main subject of this paper. The coefficient $\nu_o$ is known as kinematic odd viscosity (or Hall viscosity). In writing this term we introduced the notation
\bea
	a_i^* \equiv \epsilon_{ij}a_j
\eea
so that the ``starred'' vector $\mathbf{a}^*$ is just a vector $\mathbf{a}$ rotated by 90 degrees clockwise. 

The following three important remarks are in order. 1) All equations we wrote so far are legitimate in any spatial dimension except for the odd viscosity term of (\ref{eq:Tij}) which is specific to two-dimensional hydrodynamics. It breaks parity (explicitly uses $\epsilon_{ij}$) without breaking isotropy of the two-dimensional fluid \cite{avron1998odd}; 2) The symmetric stress tensor (\ref{eq:Tij}) does not have the most general form for two-dimensional isotropic fluids with broken parity. Within this order in velocity gradients for the incompressible fluids, one more term can be added. It is the contribution to the diagonal part of the stress proportional to $\omega \delta_{ij}$, where $\omega = \bm\nabla\times \mathbf{v}=\p_iv_i^*$ is the vorticity of the fluid (pseudoscalar in two dimensions). This term can be easily included into the discussion but is omitted for simplicity \footnote{The vorticity contribution to the pressure term of the stress tensor can fully or partially cancel the effects described in this work. The actual values of all coefficients in the  stress tensor should either be measured experimentally or derived from the underlying microscopic physics.} ; 3) For an incompressible fluid the equation of state $p(\rho)$ is not needed. The pressure $p$ in this case is not a ``state variable'' as it is fully determined by the fluid flow.

To elaborate on the third remark it is convenient to rewrite (\ref{eq:euler}) as
\bea
	D_t v_i = -\p_i(p-\nu_o \omega+gy) +\nu_e \Delta v_i \,,
 \la{eq:euler10}
\eea
where we used the expression (\ref{eq:Tij}) and the incompressibility condition (\ref{eq:incom}). Taking a curl of this equation we obtain the vorticity transport equation
\bea
	D_t \omega = \nu_e \Delta \omega \,.
 \la{eq:omegatr}
\eea
In two dimensions, the two equations (\ref{eq:incom},\ref{eq:omegatr}) define the two-component velocity field and one can find pressure $p$ from the known velocity field using (\ref{eq:euler10}). Notice that both equations (\ref{eq:incom},\ref{eq:omegatr}) do not depend directly on odd viscosity $\nu_o$. This means that the effects of odd viscosity on the flow of an incompressible fluid can come only through boundary conditions \cite{ganeshan2017odd}.

%%%%%%%%%%%%%%%%%%%%%%%%%%%%
\subsection{Boundary conditions}

Let us consider the fluid dynamics on a domain with boundary. In this case in addition to equations (\ref{eq:incom}-\ref{eq:Tij}) we need boundary conditions. In this work we are interested in the motion of the surface and, correspondingly, in free moving surface boundary conditions (for other examples, see Ref.~\cite{ganeshan2017odd}).
  %%%%%%%%%%%%%%%%%%%%%%%%%%%%%%%%%%%%%%%%%%%%%%%%%%%%%%%%%%%%%%%%%%%%%%%%%%%%%%%%%%%%%%%%
\begin{figure}
\centering
\includegraphics[scale=0.4]{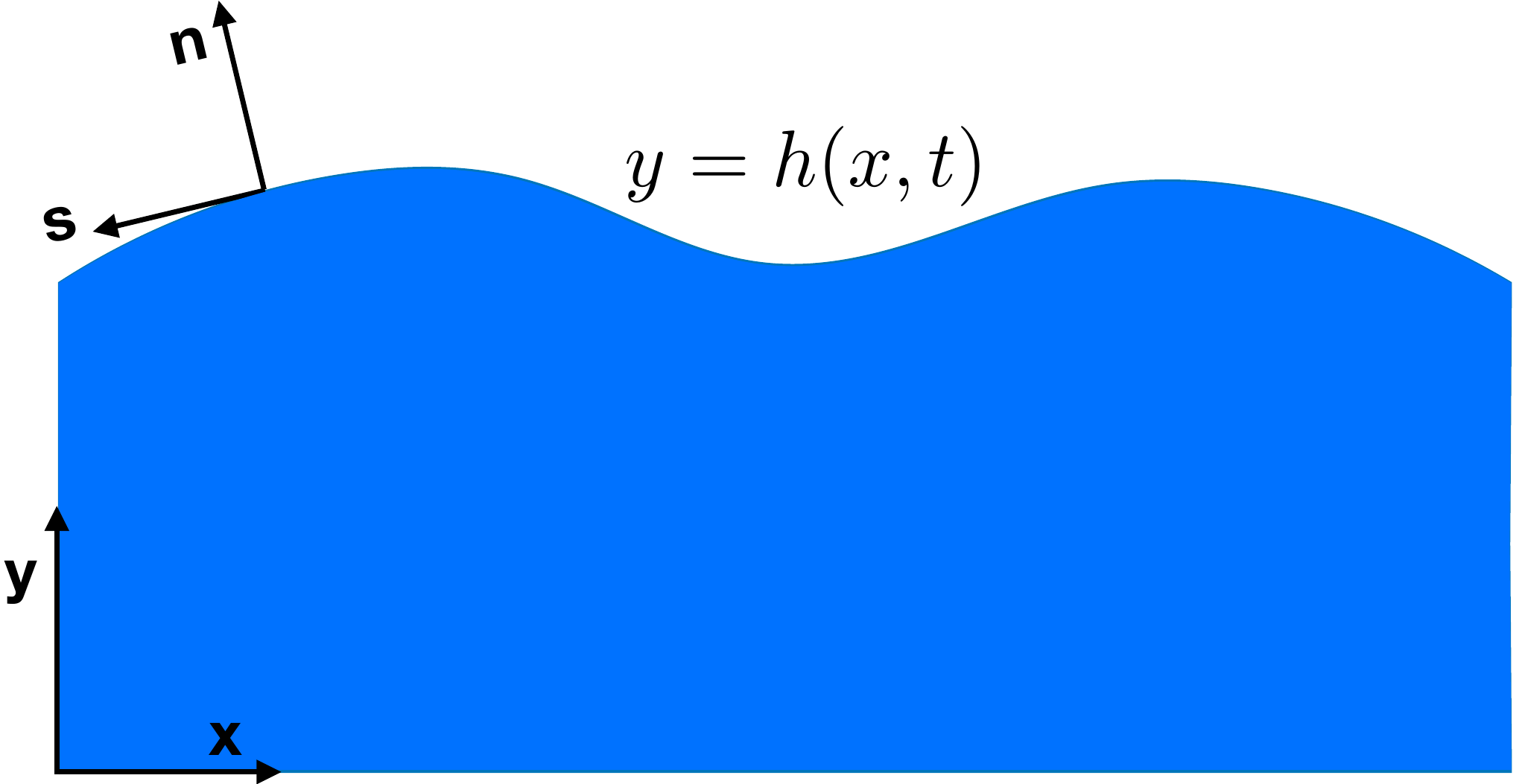}\\
\caption{\label{wave} Fluid domain with free surface $y=h(x,t)$}
\la{fig:sn}
\end{figure}
%%%%%%%%%%%%%%%%%%%%%%%%%%%%%%%%%%%%%%%%%%%%%%%%%%%%%%%%%%%%%%%%%%%%%%%%%%%%%
We parameterize the boundary of the fluid as $y=h(x,t)$ assuming that the fluid domain is defined by $y\leq h(x,t)$ (see Fig.~\ref{fig:sn}). The first boundary condition is the kinematic boundary condition (KBC) expressing the connection between the motion of the boundary of the fluid with the fluid velocity taken at the boundary 
\bea
	\partial_{t} h + v_{x} \partial_{x} h = v_{y}\,, \quad \mbox{at } y=h(x,t)\,.
 \la{eq:KBC}
\eea

In addition to KBC we have two dynamical boundary conditions (DBC) stating that the stress forces determined by (\ref{eq:Tij}) are continuous across the boundary. Here, we impose no-stress boundary conditions assuming that the fluid is in contact with a medium such as air or vacuum which cannot apply any forces except for maybe constant pressure. We have
\bea
	T_{ij}n_j= 0\,,
 \la{eq:DBC}
\eea
where the outward-pointing normal vector is given by $\mathbf{n} =\frac{1}{\sqrt{1 + (\p_x h)^{2}}}(-\p_x h, 1)$. 
%Here and in the following we use an obvious notation $h_x=\p_x h$ etc.
 
Finally, in the case when the fluid domain is non-compact one should additionally impose boundary conditions at infinity. We will assume here that those boundary conditions are $\mathbf{v}\to 0$ as $y\to -\infty$.

%%%%%%%%%%%%%%%%%%%%%%%%%%%%
%%%%%%%%%%%%%%%%%%%%%%%%%%%%
\section{Irrotational motion in the bulk}\label{sec:potflow}
%%%%%%%%%%%%%%%%%%%%%%%%%%%%

It follows from (\ref{eq:omegatr}) that an initially irrotational region of the fluid ($\omega=0$) stays irrotational until the vorticity is delivered to the region either by convection or by diffusion. Let us assume that the initial state of the fluid is irrotational. Then, as the fluid starts to move one should have vanishing vorticity for some finite time. Assuming that the motion of the fluid in the bulk is irrotational $\omega=0$ we can parameterize the velocity by a potential $v_i=\p_i\phi$ which satisfies the incompressibility condition written as the Laplace equation
\be
	\Delta \phi =0\,.
\ee
For such a potential flow, Eq.~(\ref{eq:euler10}) is equivalent to Bernoulli's equation \cite{landau1987fluid}
\bea
	\p_t\phi +\frac{1}{2}(\nabla \phi)^2 = -\tilde{p}-gy\,.
 \la{eq:Bernoulli}
\eea

One can think of (\ref{eq:Bernoulli}) as the equation determining the modified pressure 
\be
	\tilde{p}=p-\nu_o\omega
 \la{eq:modpressure}
\ee 
so that any solution of the Laplace equation depending on time as a parameter gives a local solution of the Navier-Stokes equation (\ref{eq:euler10}) for an incompressible flow. To fully specify the hydrodynamic flow one needs in addition to satisfy boundary conditions. Before discussing boundary conditions we notice that a potential $\phi(x,y,t)$ is harmonic inside the fluid domain $D$ and, therefore, is fully defined by its values $\phi|_\Sigma$ on a boundary of the domain $\Sigma =\p D$. Therefore, knowing the potential at $\Sigma$ one immediately knows the potential and velocity fields everywhere inside the domain $D$. In particular, the tangent and normal components of the velocity at the boundary of the domain $\Sigma$ can be determined from one scalar function $\phi|_\Sigma$ and are not independent from each other.

%The stress tensor (\ref{eq:Tij2}) in this case becomes {\blue Sasha: I am not sure we need it here}
%\bea
%	\frac{1}{\rho_0}T_{ij} = -\frac{p}{\rho_0}\delta_{ij} +2\nu_e \p_i\p_j\phi
%	+2\nu_o \p_i^*\p_j\phi\,.
% \la{eq:Tij3}
%\eea
%The stress force acting on a unit surface perpendicular to the unit vector $\mathbf{n}$ is given by
%\bea
%	f_i = T_{ij}n_j = -pn_i +\rho_0 \p_n\Big(2\nu_e \p_i\phi
%	+2\nu_o \p_i^*\phi\Big)\,.
% \la{eq:fi}
%\eea
%where $\p_n=n_j\p_j$ is a derivative in the direction of $\mathbf{n}$. 

In the absence of viscosities, DBC (\ref{eq:DBC}) reduces to a single requirement $p|_\Sigma=0$ which can be satisfied by the proper choice of $\phi|_\Sigma$. On the other hand, when one or both viscosity coefficients $\nu_{e,o}$ are non-vanishing, the free surface DBC (\ref{eq:DBC}) produce two non-trivial boundary conditions. These conditions are impossible to satisfy just by fine tuning the scalar potential $\phi\Big|_\Sigma$.  

A resolution of this problem is well known in the absence of odd viscosity but with non-zero shear viscosity $\nu_e$. The irrotationality assumption on the fluid flow should necessarily break down at the moving free surface. A time-dependent boundary layer with non-vanishing vorticity is formed at the free surface of the moving fluid \cite{batchelor2000introduction,lamb1932hydrodynamics}. In this paper, we study the effects of the boundary layer on the motion of the surface in the presence of odd viscosity. Before going into this, let us first study in some detail the solution of the linearized problem applicable in the case of small amplitude surface waves.

%%%%%%%%%%%%%%%%%%%%%%%%%%%%
%%%%%%%%%%%%%%%%%%%%%%%%%%%%
\section{Linear odd surface waves}
 \la{sec:linoddwaves}
%%%%%%%%%%%%%%%%%%%%%%%%%%%%

Modifying the derivation of Lamb~\cite{lamb1932hydrodynamics} for the case of non-vanishing odd viscosity, we derive a solution corresponding to surface waves in the presence of both $\nu_e$ and $\nu_o$ in the limit of small amplitude oscillations.
We start by linearizing the basic equations (\ref{eq:euler10}) assuming that velocity, its gradients as well as the gradients of external potential are small. We obtain the following linear bulk equations:
\bea
	{\bf \nabla}\cdot \mathbf{v} &=& 0\,,
 \la{eq:incom1} \\
	\p_t \mathbf{v} &=& -{\bf \nabla}\tilde{p} +\nu_e \Delta \mathbf{v} -g \hat{\mathbf{y}}\,.
 \la{eq:euler10lin} 
\eea
These equations are identical to the equations for the fluid with $\nu_o=0$ (with the only change $p\to\tilde p=p-\nu_{o}\omega$ and we can borrow the Lamb's solution for plane waves written using complex valued velocities (see \cite{lamb1932hydrodynamics}, art.~349)
\begin{align}
	v_{x} &=  \left( A |k| e^{ |k| y} + B m e^{ my}\right)e^{ i k x - i \Omega t} \,,
 \label{eq:linvx}\\
	v_{y} &= -ik \left(A e^{ |k| y} +B e^{ m y}\right)e^{ i k x - i \Omega t} \,,
 \label{eq:linvy}\\
 	\omega &= e^{ i k x - i \Omega t}\left(  \frac{i \Omega}{\nu_{e} } B e^{ m y}\right) \,,  
 \label{eq:linvort} \\
	\tilde{p} &=  \Omega \frac{k}{|k|} A e^{ |k| y} e^{ i k x - i \Omega t} -gy\,.
 \label{eq:pressure}
\end{align}
Here we use the convention that the actual solution is given by the real parts of these formulas.
Here $A,B$ are arbitrary (complex) amplitudes and the following relation between $m$ and $k$ should be satisfied:
\bea
	m^{2} = k^{2} - \frac{i \Omega}{\nu_{e}}\,.
 \la{eq:mkOmega}
\eea
Here we assume that $\R(m)>0$ so that the velocity vanishes in the limit $y\to-\infty$. We also linearize the DBC (\ref{eq:DBC}) taking the normal vector $\mathbf{n}\approx (0,1)$
\begin{align}
 	p=\nu_{e}(\partial_{y} v_{y}-\partial_{x} v_{x}) - \nu_{o}(\partial_{x} v_{y} + \partial_{y} v_{x})\,,
 \label{eq:lineardbcN} \\
	\nu_{e} (\partial_{x} v_{y} + \partial_{y} v_{x}) + \nu_{o}(\partial_{y} v_{y}-\partial_{x} v_{x}) =  0\,,
 \label{eq:lineardbcT}
\end{align}
and also the KBC (\ref{eq:KBC})
\bea
	v_y=\p_t h\,.
 \la{eq:linearkbc}
\eea
%
%
% We consider the fluid to be incompressible $C=0$. The linearized equations of motion and the boundary conditions for the above set of equations can be written in the vector form as,
%\begin{align}
%\partial_{t} {\bf v} = - \nabla p + \nu_{o} \Delta {\bf v}^{*} +\nu_{e} \Delta {\bf v},\quad \bf \nabla. {\bf v}=0 
%\label{eq:lineareommain}
% \end{align}
% Using the incompressibility condition, we can absorb the odd viscosity term by using a modified pressure term $\tilde p=p-\nu_o \omega$. In terms of the modified pressure the equation of motion can be written as,
%\begin{align}
%\partial_{t} {\bf v} = - \nabla \tilde p +\nu_{e} \Delta {\bf v}.\label{eq:lineartildep}
% \end{align}	
%In addition to these linearized equations of motion, we have two dynamical boundary conditions given by,
%\begin{align}
% \tilde p=2\nu_{e}\partial_{y} v_{y} - 2\nu_{o}\partial_{x} v_{y},\\
%\nu_e(\omega+2\partial_{y} v_{x})=2\nu_{o}\partial_{x} v_{x}.
%\label{eq:lineardbclamb}
%\end{align}
%Notice that the DBC from the normal stress tensor tensor fixes the pressure at the surface. We have expressed the tangential stress condition tensor using vorticity $\omega$ at the boundary. Satisfying tangential stress conditions at the boundary is equivalent to fixing vorticity at the boundary. The kinematic boundary condition given by $v_y=\p_t h(x,t)$.
%
Substituting (\ref{eq:linvy}) into (\ref{eq:linearkbc}) we obtain
\begin{align}
	h (x, t) &=  \frac{k}{\Omega}\left(A+B\right)e^{ i k x - i \Omega t} \,.
 \la{eq:linh}
\end{align}
We note here that within the linear approximation one can evaluate fields at $y=0$ instead of $y=h(x,t)$. The only exception is the last term of (\ref{eq:pressure}) which should be replaced by $-gh$ at the boundary. Substituting (\ref{eq:linvx},\ref{eq:linvy},\ref{eq:pressure},\ref{eq:linh}) into the DBC (\ref{eq:lineardbcN},\ref{eq:lineardbcT}) we obtain two equations for the amplitudes
\bea
	A\Big[g|k|-\Omega^{2}-2\Omega(\nu_{o}k|k| + i\nu_{e}k^{2})\Big]
	+B\Big[g|k|-2\Omega (\nu_{o}k|k| + i\nu_{e}|k| m)\Big] 
	&=& 0\,,
 \label{eq:AB1} \\
	A\Big[2(\nu_{o}k|k|+i\nu_{e}k^{2})\Big]
	+B\Big[\Omega +2\nu_{o}km+2i\nu_{e}k^{2}\Big] 
	&=& 0\,.
 \label{eq:AB2}
\eea

Equations (\ref{eq:AB1},\ref{eq:AB2}) together with the relation (\ref{eq:mkOmega}) fully define the solution for surface waves. In particular, the dispersion relation $\Omega(k)$ is obtained by requiring consistency of (\ref{eq:AB1}) and (\ref{eq:AB2}). The full expression for $\Omega(k)$ is not very illuminating. We analyze the dispersion relation and its different limits in Appendix~\ref{app:dispersion}.

%%%%%%%%%%%%%%%%%%%%%%%%%%%%
\subsection{Gravity waves in the presence of small odd viscosity}
 \la{sec:gwson}

In a limit of small odd and even viscosities we obtain a perturbative correction to Lamb's solution (see Appendix~\ref{app:dispersion})
\begin{align}
	\Omega & \approx \pm  \sqrt{g |k|} - 2 i \nu_{e} k^{2} - 2 \nu_{o} k |k| \,.
 \la{eq:Omegagn}
\end{align}
In the limit $\nu_o=0$ the dispersion (\ref{eq:Omegagn}) reduces to the well-known result for incompressible waves with shear viscosity \cite{lamb1932hydrodynamics}. 

%%%%%%%
\begin{figure}
\centering
\includegraphics[scale=0.6]{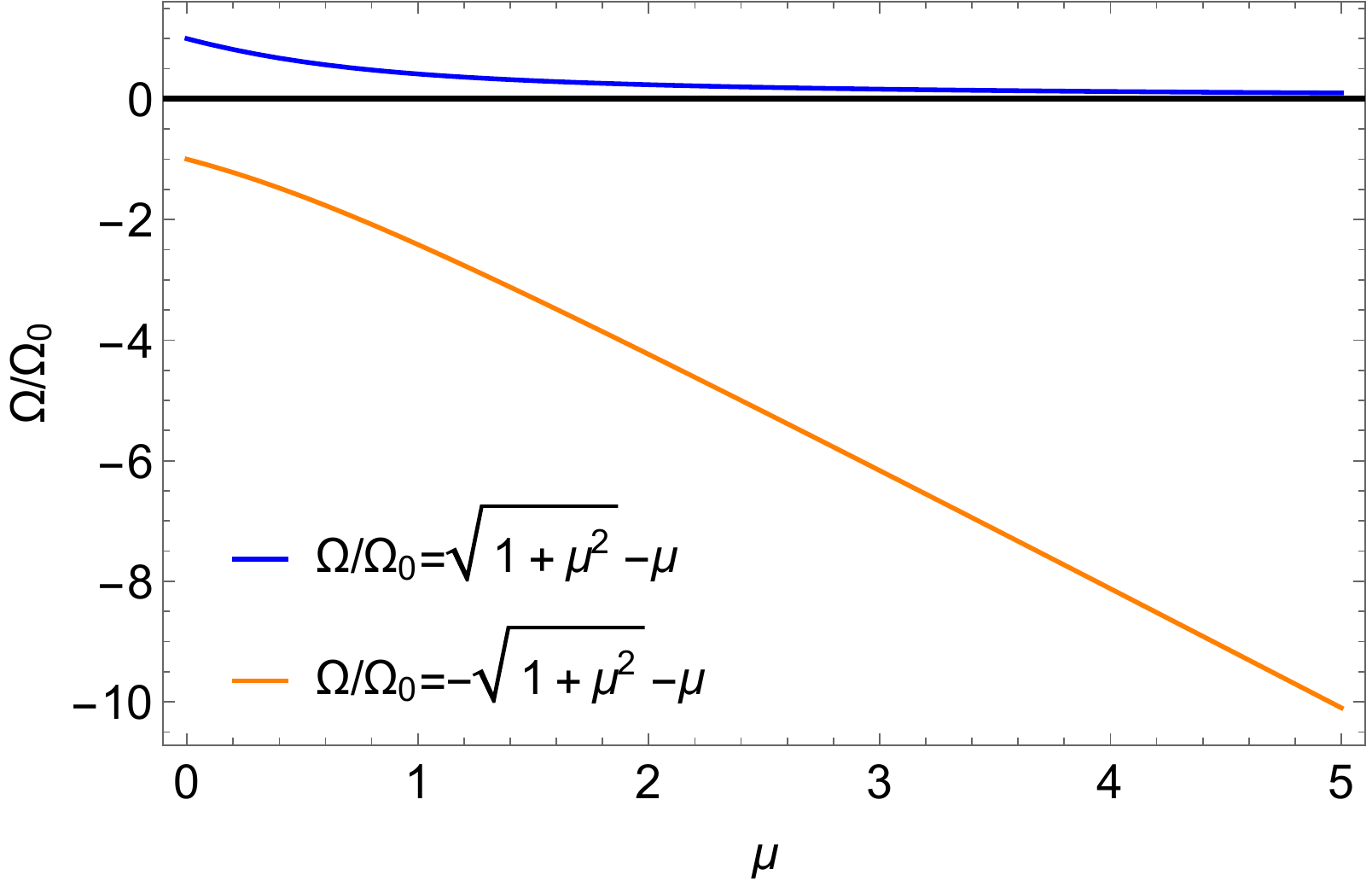}\\
\caption{ The dispersion of surface waves as a function of odd viscosity at vanishing shear viscosity $\nu_e\to 0$. Here $\Omega_0=\sqrt{gk}$ and $\mu=\nu_o k^2/\Omega_0$ is a dimensionless odd viscosity.}
\label{fig:omega}
\end{figure}
%%%%%%%

It is clear from the relation (\ref{eq:mkOmega}) that there is a typical length scale $\delta = \sqrt{\nu_e/|\Omega|}$. As the frequency $\Omega$ of the surface waves remains finite in the limit $\nu_e\to 0$ we find that the $B$-component of the solution (\ref{eq:linvx},\ref{eq:linvy}) and vorticity of the fluid (\ref{eq:linvort}), in particular, is localized near the surface of the fluid within the dynamic boundary layer of thickness $\delta$. For conventional slightly damped gravity waves (without odd viscosity) $\Omega\approx\pm \sqrt{g|k|}$ and the existence and structure of such a layer is well known \cite{lamb1932hydrodynamics,batchelor2000introduction}. In the following we will see how this boundary layer is modified by odd viscosity.

%%%%%%%%%%%%%%%%%%%%%%%%%%%%
\subsection{Surface waves dominated by odd viscosity in the absence of gravity}
 \la{sec:swodd}

In the following we consider the novel regime when surface waves are dominated by the odd viscosity and the gravity is absent. Namely, we assume that $\nu_o\gg\nu_{e}$ and switch off the external potential $g=0$. The equation for the wave dispersion in the absence of gravity (see Appendix~\ref{app:dispersion}) is:
\bea
 \Omega^{3} +2\Omega^2 k\Big(\nu_o(m+|k|)+2i \nu_e k\Big) 
	+ 4\Omega k^{2}|k| (m-|k|)(\nu_{o}^{2}+\nu_{e}^{2}) = 0\,.
\eea

There are only two solutions of this equation that correspond to the wave decaying into the bulk of the fluid (i.e., $\re(m)>0$, see Fig.~\ref{fig:omega}). The first one is trivial, corresponding to the arbitrary deformation of the boundary of the fluid $y=h(x)$ which is at rest: $\Omega=0$, $\mathbf{v}=0$. In the absence of gravity, there is no restoring force and the fluid remains motionless (we neglect surface tension in this work). However, in the presence of odd viscosity there exists also a non-trivial solution 
\begin{align}
	\Omega \approx -2\nu_o k |k| - i \sqrt{\nu_e|\nu_o|} k^2\,,
 \label{eq:dispmain}
\end{align}
where we keep only the leading terms in real and imaginary parts of $\Omega(k)$. The first term of (\ref{eq:dispmain}) corresponds to a chiral wave propagating with group velocity 
\be
	v_k = \frac{\p \Omega}{\p k} \approx -4\nu_o |k| \,.
 \la{eq:groupv}
\ee
The direction of the propagation of the wave is determined by the sign of the odd viscosity $\nu_o$. The second term (\ref{eq:dispmain}) describes the damping of the wave. The details of the derivation of the linear wave solution and its dispersion are given in Appendix~\ref{app:dispersion}. The limit of $\nu_e\to 0$ in (\ref{eq:mkOmega},\ref{eq:AB1},\ref{eq:AB2}) produces in the leading non-vanishing order
\bea
	m &\approx & (1+i\sign(\nu_o k))\sqrt{\frac{|\nu_o|}{\nu_e}}\, |k| \,,
 \la{eq:mlin}\\
 	A &\approx & -2\nu_o |k| D \,,
 \\
 	B &\approx & -A \frac{1-i\sign(\nu_o k)}{2} \sqrt{\frac{\nu_e}{|\nu_o|}} \,.
 \la{eq:BA}
\eea
The height profile and the vorticity is given by,
\bea
	h &\approx & D e^{ikx-i\Omega t},\\
	\omega & = & -2\nu_o k^2|k| D (1+i\sign(\nu_o k))\sqrt{\frac{|\nu_o|}{\nu_e}} 
	e^{my+ikx-i\Omega t}. 
\eea
and in the regime $\nu_e\ll \nu_o$ the vorticity is non-vanishing only in the narrow layer of the thickness of the order of $\delta=k^{-1}\sqrt{\frac{\nu_e}{|\nu_o|}}$ which can be found as $m^{-1}$ from (\ref{eq:mlin}). We emphasize that the diverging of vorticity as $\omega\sim1/\sqrt{\nu_e}$ in the boundary layer is solely due the the presence of odd viscosity and is dramatically different compared to the Lamb's case where $\omega\sim O(1)$ as $\nu_e\rightarrow 0$. In the limit of interest $B\ll A$ (\ref{eq:BA}) and the vertical component of the velocity is essentially defined by the $A$-component of the solution 
\be
	v_y \approx -ik A e^{|k|y} e^{ikx-i\Omega t}\,. 
\ee
On the contrary the horizontal component of the velocity has contributions from both $A$ and $B$ parts of the solution. We obtain
\be
	v_x \approx A|k| \left(e^{|k|y}-e^{my}\right) e^{ikx-i\Omega t} \,.
 \la{eq:horv}
\ee
At the surface $y\approx 0$ the horizontal velocity is $\sim \sqrt{\nu_e}$ and vanishes within our accuracy (\ref{eq:horv}). However, it changes rapidly with $y$ and becomes of the order of $ A|k|$ at the depth of few $\delta$. This is the $v_x$ which contributes the most to vorticity so that near the surface the vorticity diverges $\omega \approx -\p_y v_x \sim \nu_e^{-1/2}$ while the $v_x$ changes by a finite amount ($\sim \nu_e^0$) across the layer of  thickness $\delta \sim \nu_e^{1/2}$. This behavior is rather different from Lamb's solution, in which only the vortical part (B-part) of $v_x$ vanishes as $\nu_e\rightarrow 0$. 

%%%%%%%%%%%%%%%%%%%%%%%%%%%%%%%%%%%%%%%%%%%%%%%%%%%%%%%%%%%%%%%%%%%%%%%%%%%%%%%%%%%%%%%%
\begin{figure}
\centering
\includegraphics[scale=0.4]{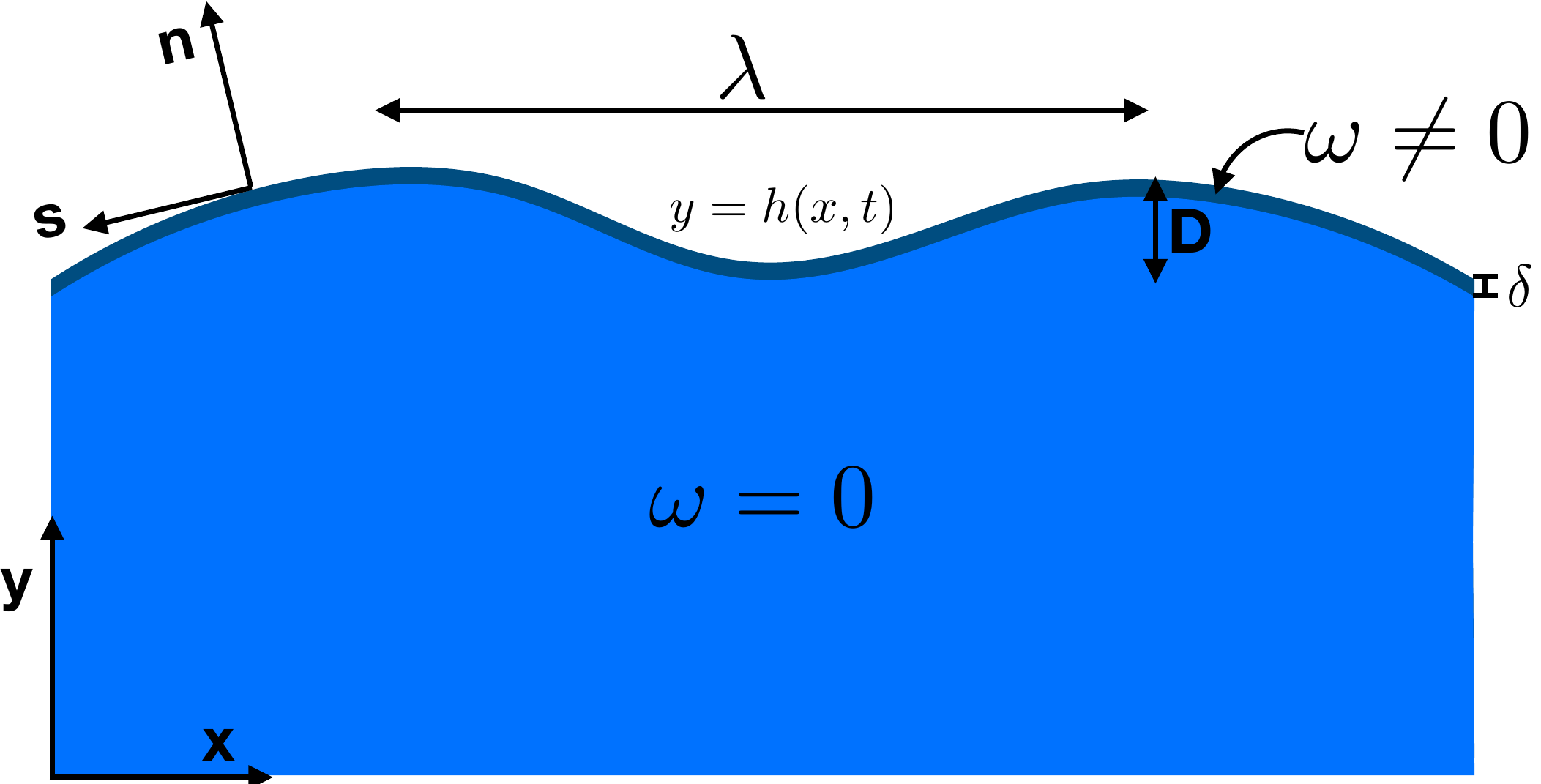}\\
\caption{\label{fig:bdylayer} Surface waves with boundary layer. }
\end{figure}
%%%%%%%%%%%%%%%%%%%%%%%%%%%%%%%%%%%%%%%%%%%%%%%%%%

%%%%%%%%%%%%%%%%%%%%%%
\section{Effective nonlinear boundary dynamics}
 \la{sec:effheur}
%%%%%%%%%%%%%%%%%%%%%%

In this section we present physical arguments leading to the effective nonlinear equation for surface waves relegating more precise arguments and detailed derivation to Appendix~\ref{app:QPA}. Let us consider a nonlinear wave in the limit of small $\nu_e$ with the amplitude $D$ small compared to the wavelength of the wave $\lambda$ but non-vanishing (weakly nonlinear waves). Here we also assume that the external gravity field is absent.  In Section~\ref{sec:linoddwaves} we saw that in this limit the typical frequency $\Omega\sim \nu_o k^2$ remains finite and that the vorticity is confined to the narrow layer near the surface. Motivated by the structure of the linear solution we make the following assumptions for weakly nonlinear waves:
\begin{enumerate}
	\item A typical time scale (frequency) remains finite
	in the limit of vanishing shear viscosity $\nu_e$.
	\item The motion is potential (irrotational) in the bulk with 
	vorticity localized in the narrow region near the surface --- the boundary layer 
	(see Fig.~\ref{fig:bdylayer}). The thickness of this region is much 
	smaller than the amplitude of the wave $D$.
\end{enumerate}

The main idea of the following derivation is to consider the effect of the narrow boundary layer on the potential flow of the fluid outside of the boundary layer. We essentially ``integrate out'' the layer and consider the potential flow with modified boundary conditions for the rest of the fluid.\footnote{In contrast to this approach in the more precise approach of Appendix~\ref{app:QPA} we decompose velocity into potential and vortical part and integrate out the vortical part, again obtaining the effective equations for potential flow.} First, let us consider the boundary layer.

Neglecting the curvature of the surface, we see from the tangent stress boundary condition (\ref{eq:lineardbcT}) that in the limit $\nu_o\gg \nu_e$ we essentially have $\p_yv_y-\p_xv_x\approx 0$ (or $\p_xv_x\approx 0$ using incompressibility). This means that for small curvature $D/\lambda \ll 1$ the tangent component of the fluid velocity should vanish at the free surface\footnote{More precisely, the tangent velocity at the boundary is of the second order in amplitude. For detailed scaling analysis, see Appendix.~\ref{app:QPA}}. The odd viscosity imposes an effective no-slip boundary condition for the tangential velocity on the moving surface. The finite curvature generates a tangent velocity at the boundary but it remains small and does not invalidate the boundary layer approximation (see Appendix \ref{app:QPA}). Having small tangent velocity at the boundary and finite velocity right below the boundary layer means that large vorticity is generated inside the layer (c.f., Section \ref{sec:swodd}). The small but finite shear viscosity $\nu_e$ is important in determining the thickness of this layer but as we will see drops out of the effective equation for velocity potential. In the limit $\nu_e\to 0$ the boundary layer is infinitesimally thin and we can 
write the KBC (\ref{eq:KBC}) as
\bea
	h_t =v_y-h_xv_x = \p_y\phi-h_x\p_x\phi\,.
 \la{eq:KBCodd}
\eea
Here the velocity $v$ is taken at the free surface and the potential $\phi$ is taken right below the boundary layer. As $v_x$ is already of the second order of smallness in the amplitude, we can drop the term $h_xv_x$ in quadratic approximation and obtain the relation between velocity at the free surface and potential as 
\be
	v_y \approx \p_y\phi-h_x\p_x\phi\,.
 \la{eq:vy}
\ee
Let us now consider the normal stress boundary condition. In the limit of $\nu_e\to 0$ we can write the stress tensor (\ref{eq:Tij}) as
\bea
	T_{ij} = -\tilde{p}\delta_{ij}+2\nu_o \p_i^*v_j\,.
\eea
The normal stress boundary condition at the free surface becomes
\bea
	0=T_{nn} = -\tilde{p} +2\nu_o n_in_j \p_i^* v_j \approx -\tilde{p}-2\nu_o(\p_xv_y-2h_x\p_xv_x)\,,
\eea
where we used $\hat{n}\approx (-h_x,1)$ and kept only terms up to quadratic ones in $h$ and $v$. As the tangent velocity at the boundary is already of the second order in the amplitude, the term $h_x\p_x v_x$ can be neglected in quadratic approximation. Using (\ref{eq:vy}) we obtain
\be
	\tilde{p} = -2\nu_o\p_x v_y = -2\nu_o \p_x(\p_y\phi-h_x\p_x\phi)
 \la{eq:tildep}
\ee
as an effective dynamical boundary condition. The modified pressure does not change across the very thin boundary layer and we can use (\ref{eq:tildep}) as the boundary condition just below the boundary layer.

Now consider the Bernoulli equation (\ref{eq:Bernoulli}) which is expected to be valid right below the boundary layer. Substituting the expression of modified pressure (\ref{eq:tildep}) and with $g=0$ we have:
\bea
	\p_t\phi +\frac{1}{2}(\nabla \phi)^2 =2\nu_o \p_x(\p_y\phi-h_x\p_x\phi)\,.
 \la{eq:Bernoulli2}
\eea
This equation is assumed to be valid at the surface $y=h(x,t)$, where we neglect the thickness of boundary layer and essentially replace the true physical surface of the fluid by the inner surface of the boundary layer. 
%Notice that $v_x$ is very small at the surface of the fluid and the KBC (\ref{eq:KBC}) can be written as
%\bea
%	\p_t h =\p_y\phi\,,
% \la{eq:KBCodd}
%\eea
%where we expressed the normal component of velocity in terms of potential assuming that the normal component of velocity does not change much across the very thin boundary layer. 
To proceed we introduce the boundary value of the potential
$$
	\tilde\phi(x,t) = \phi(x,h(x,t),t)\,.
$$
We find in quadratic approximation
$$
	\p_t\tilde\phi = \p_t\phi +h_t\p_y\phi \approx \p_t\phi +(\p_y\phi)^2.
$$
and substituting into (\ref{eq:Bernoulli2})
\begin{align}
	\p_t\tilde \phi +\frac{1}{2}(\p_x\phi)^2-\frac{1}{2}(\p_y\phi)^2 
	=2\nu_o\p_x(\p_y\phi-h_x\p_x\phi)\,.
 \la{eq:Bernoulli4}
\end{align}
Finally, we remember that the potential $\phi$ is harmonic in the lower half plane (strictly speaking, it is harmonic for $y\leq h(x,t)$). Rewriting the above equation in terms of $\tilde \phi(x,t)$ (see Eq.~\ref{eq:phiycomb} and Appendix~\ref{app:hilbert} for details) and keeping up to quadratic terms we obtain,
\be
	\tilde\phi_t+\frac{1}{2}( \tilde\phi_x^2 -(\tilde\phi_x^H)^2)+2\nu_o\tilde\phi^H_{xx} 
	= -2\nu_o\Big[h\tilde\phi_{x}+(h\tilde\phi_x^H)^H\Big]_{xx} \,.
 \la{eq:Bernoulli3}
\ee
Here ``$H$'' denotes Hilbert transform \footnote{\la{footnote-hilbert}  If the function $\phi(x,y)$ is harmonic in the lower half plane $y\leq 0$, the complex function $2\phi^-\equiv \phi+i\phi^H$ is analytic in the lower half plane of $z=x+iy$. Its real and imaginary parts satisfy the Cauchy-Riemann relations $\p_x\phi=\p_y\phi^H$ and $\p_y\phi=-\p_x\phi^H$. The relations used in the text are derived in Appendix~\ref{app:hilbert}. Appendix includes this derivation to next order correction in nonlinearity. These corrections are due to the difference between $\phi(x,0)$ and $\phi(x, h(x,t)) \equiv \tilde\phi(x,t)$ and between the true domain of harmonicity $y\leq h(x)$ and the lower half plane $y\leq 0$.}
 defined as
\be
	\phi^H(x) = P.V. \frac{1}{\pi}\int_{-\infty}^{+\infty}\frac{\phi(x')dx'}{x'-x}\,.
 \la{eq:Htransf}
\ee
The KBC (\ref{eq:KBCodd}) becomes
\begin{align}
	h_t +\tilde\phi_x^H=-\Big[h\tilde\phi_x  +(h\tilde\phi_x^H)^H\Big]_x  \,.
 \la{eq:ht1}
\end{align}
This form does not depend on $\nu_o$ and coincides with Kuznetsov et al. \cite{kuznetsov1994formation}. In terms of $\tilde{u}=\tilde\phi_x$ it becomes
\begin{align}
	h_t +\tilde{u}^H+\Big[h\tilde{u} +(h\tilde{u}^H )^H\Big]_x \approx 0 \,.
\end{align}
Differentiating (\ref{eq:Bernoulli3}) with respect to $x$ and introducing $\tilde u=\tilde\phi_x$ we obtain from (\ref{eq:Bernoulli3},\ref{eq:ht1}) the following system of equations:
\begin{align}
	\tilde u_t +\tilde u\tilde{u}_x- \tilde{u}^H \tilde{u}^H_x +2\nu_o\tilde{u}^H_{xx} 
	&= -2\nu_o\Big[h\tilde{u}+(h\tilde{u}^H)^H\Big]_{xxx} \,,
 \la{eq:Bernoulli5} \\
 	h_t +\tilde{u}^H
	&=-\Big[h\tilde{u}  +(h\tilde{u}^H)^H\Big]_x  \,.
 \la{eq:ht2}
\end{align}
This nonlinear system is Hamiltonian! It is remarkable that the Hamiltonian itself is the same as for the fluid without odd viscosity and is equal to the total kinetic energy of the fluid. However, the Poisson brackets of the fields are modified by odd viscosity dependent terms. For details see next section.

In the small surface angle approximation (see Ref.~ \cite{kuznetsov1994formation}) we neglect the right hand side of (\ref{eq:Bernoulli5}) and (\ref{eq:ht2}) and obtain
\begin{align}
	\tilde u_t +\tilde u\tilde{u}_x- \tilde{u}^H \tilde{u}^H_x +2\nu_o\tilde{u}^H_{xx} 
	&= 0\,,
\la{eq:almostBO1} \\
 	h_t +\tilde{u}^H
	&=0  \,.
 \la{eq:ht3}
\end{align}
The first equation reminds us of the well-known Benjamin-Davis-Ono (BDO)\footnote{The Benjamin-Davis-Ono equation has a form $u_t+uu_x+u^H_{xx}=0$.}  equation but the nonlinear term is very different. This equation results in a new class of non-linear chiral dynamics which we dub as chiral Burgers equation  (complex Burgers equation with an additional analyticity condition, see section \ref{sec:nln}).
We notice here that linearizing (\ref{eq:Bernoulli5},\ref{eq:ht2}) and including gravity we obtain
\begin{align}
	\tilde u_t +2\nu_o\tilde{u}_{xx}^H+gh_x
	&= 0\,,
\la{eq:almostBO1-lin} \\
 	h_t +\tilde{u}^H
	&=0  \,.
 \la{eq:ht3-lin}
\end{align}
It is easy to check that these linear equations produce the dispersion given by (\ref{eq:Omegamu}). Also, in the absence of gravity $\tilde{u}_t=2\nu_o (h_{xx})_t$ so that the tangent velocity at the bottom of the boundary layer changes in proportion to the curvature of the surface $-h_{xx}$.

To summarize, the tangent stress boundary condition is satisfied through the formation of the vortical boundary layer at the surface of the fluid. Right below the boundary layer the modified pressure is given by $\tilde{p}\approx -2\nu_o\p_x v_y $. Using the potentiality of the flow in the bulk below the boundary layer we arrived at the Hamiltonian system (\ref{eq:Bernoulli5},\ref{eq:ht2}). An additional small surface angle approximation gives a simplified system\footnote{Unfortunately, the simplified system (\ref{eq:almostBO1},\ref{eq:ht3}) is not Hamiltonian.} (\ref{eq:almostBO1},\ref{eq:ht3}).

%%%%%%%%%%%%%%%%%%%%%%
\section{Hamiltonian structure}
 \la{sec:hamiltonian}
%%%%%%%%%%%%%%%%%%%%%%
As described in the previous section, in the limit $\nu_e\to 0$ the dissipation vanishes and we expect the effective surface dynamics equations to have a Hamiltonian structure. In this section, we show that this is indeed so. Let us first consider the total kinetic energy of the fluid\footnote{We neglect the kinetic energy of the vortical layer as it is vanishing in the limit $\nu_e\to 0$.}
\begin{align}
	E = \int dx\int_{-\infty}^{h(x)} dy\,\frac{1}{2}(\phi_x^2+\phi_y^2)
	= \int dx\int_{-\infty}^{h(x)} dy\,\frac{1}{2}\Big((\phi\phi_x)_x+(\phi\phi_y)_y\Big)\,.
 \nonumber 
\end{align}
Here we used the incompressibility condition in the bulk $\phi_{xx}+\phi_{yy}=0$. We proceed by applying Green's theorem and using (\ref{eq:phiycomb}) 
\begin{align}
	E &= \int dx\,\left\{-\frac{1}{2}h_x\phi\phi_x+\frac{1}{2}\phi\phi_y\right\}_{y=h}
	= \int dx\,\frac{1}{2}\tilde\phi(\p_y\phi-h_x\p_x\phi)_{y=h}
 \nonumber \\
 	&\approx \int dx\,\frac{1}{2}\tilde\phi\Big(-\tilde\phi_x^H - (h\tilde\phi_{x})_x -(h\tilde\phi_x^H)_x^H\Big)
	=\int dx\,\left\{\frac{1}{2}h\left[(\tilde\phi_x)^2-(\tilde\phi_x^H)^2\right]
	-\frac{1}{2}\tilde\phi\tilde\phi_x^H\right\}\,.
 \la{eq:E}
\end{align}
The obtained expression coincides with the one of Ref.~\cite{kuznetsov1994formation}.
Let us now consider a Hamiltonian which is given by the expression (\ref{eq:E}) with the addition of gravitational potential energy
\begin{align}
	H = \int dx\,\left\{ h \Big[\frac{1}{2}(\tilde\phi_x)^2-\frac{1}{2}(\tilde\phi_x^H)^2\Big]
    -\frac{1}{2}\tilde\phi\tilde\phi_x^H
    +\frac{1}{2}gh^2
    \right\}\,.
 \la{eq:Ham1}
\end{align}
We can easily calculate the variation of the Hamiltonian (\ref{eq:Ham1}) to obtain
\begin{align}
	\delta H &= \int dx\, \left\{\delta h \Big[\frac{1}{2}(\tilde\phi_x)^2-\frac{1}{2}(\tilde\phi_x^H)^2+gh\Big]
 	- \delta\tilde\phi \Big[\tilde\phi^H+h\tilde\phi_x+(h\tilde\phi_x^H)^H\Big]_x \right\}
    \,.
 \la{eq:Hamvar}
\end{align}
We also consider slightly non-canonical Poisson brackets 
\begin{align}
	\{\tilde\phi,h'\} &= \delta(x-x')\,,
 \la{eq:PBphih} \\
 	\{\tilde\phi,\tilde\phi'\} &= -2\nu_o\p_x\delta(x-x')\,,
 \la{eq:PBphiphi}\\
 	\{h,h'\} &= 0\,.
 \la{eq:PBhh}
\end{align}
 Instead of verifying Jacobi identity for brackets (\ref{eq:PBphih}-\ref{eq:PBhh}) we notice that defining
\begin{align}
	\Phi = \tilde\phi-\nu_o h_x
 \la{eq:quasiP} 	
\end{align}
we obtain canonical Poisson brackets between $h$ and $\Phi$. 

The equations of motion generated by the Hamiltonian $H$ with the use of these Poisson brackets are
\begin{align}
	\tilde\phi_t +\frac{\delta H}{\delta h} &= 2\nu_o\p_x\frac{\delta H}{\delta\tilde\phi}\,,\quad h_t = \frac{\delta H}{\delta\tilde\phi}\,.
 \la{eq:PBeqm}
\end{align}
The equations of motion (\ref{eq:PBeqm}) from the above variational formula (\ref{eq:Hamvar}) reproduces (\ref{eq:Bernoulli5},\ref{eq:ht2}) with $g$. Using (\ref{eq:ht2}) we can write the first equation (\ref{eq:Bernoulli5}) in a more compact form
\begin{align}
	\tilde\phi_t &+\frac{1}{2}( \tilde\phi_x^2 -(\tilde\phi_x^H)^2) +g h
    =2\nu_o h_{xt} \,.
 \la{eq:tu2} 
\end{align}

%%%%%%%%%%%%%%%%%%%%%%
\subsection{Conservation laws}
In this section, we demonstrated the existence of a Hamiltonian structure to the non-linear dynamics up to the quadratic order. This automatically implies the integral of motion of (\ref{eq:Bernoulli5},\ref{eq:ht2}) given by (\ref{eq:Ham1}). In addition to energy, it is easy to check that the following quantities are conserved by (\ref{eq:Bernoulli5},\ref{eq:ht2})
\begin{align}
	M &= \int_{-\infty}^{\infty} dx\, h\,,
 \la{eq:masscon} \\
 	P &= \int_{-\infty}^{\infty} dx\, h(\tilde\phi_x-\nu_0 h_{xx}) =\int_{-\infty}^{\infty} dx\, h\Phi_x\,.
 \la{eq:momentumcon}
\end{align}
The first one is easily recognized as a conservation of fluid mass. The conservation law (\ref{eq:momentumcon}) is a consequence of translational invariance of equations (\ref{eq:Bernoulli5},\ref{eq:ht2}). Indeed, remembering the canonical Poisson brackets of $h$ and $\Phi$ in Eq.~\ref{eq:quasiP} one can easily recognize in (\ref{eq:momentumcon}) the generator of translations in the $x$ direction. The quantity $P$ appearing in the effective one-dimensional equations is related to the pseudomomentum of the two-dimensional fluid. The pseudomomentum conservation arises when the wave disturbance is translated without translating the medium in which it exists. This is a symmetry only when both the background space and the medium are homogeneous. \cite{mcintyre1981wave, stone2002phonons, falkovich2011fluid}. 

%%%%%%%%%%%%%%%%%%%%%%
\section{Nonlinear surface waves: Chiral Burgers equation}
 \la{sec:nln}
%%%%%%%%%%%%%%%%%%%%%%

We now show that the small angle dynamics of the parent Hamiltonian system (\ref{eq:almostBO1}) results in what we call chiral Burgers equation. We introduce $\tilde{u}=u(x)+\bar{u}(x)$, where $u(z)$ is the function analytic in the lower half-plane of $z$ and $\bar{u}$ its complex conjugate on real axis. Then the Hilbert transform (\ref{eq:Htransf}) is equivalent to $\tilde{u}^H=i(\bar{u}-u)$. 
The equation (\ref{eq:almostBO1}) can be decoupled into the equations for the parts holomorphic in the lower and upper half-planes. In particular for the part holomorphic in the lower half-plane we obtain (\ref{eq:almostBO}) which we reproduce here for the reader's convenience
\begin{align}
	u_t + 2uu_x - 2i\nu_o u_{xx} = 0\,.
  \label{eq:almostBO-duplicate}
\end{align}
This is a complex Burgers equation with imaginary viscosity, i.e. coefficient in front of the Burgers term $u_{xx}$. The complex Burgers equation is a simple example of a nonlinear dispersive equation. It was described in \cite{dobrokhotov1992problem} with no physical connections. In addition to (\ref{eq:almostBO-duplicate}) we have an additional analytic constraint on the solution $u(x,t)$. The function $u(x,t)$ is subject to an additional condition that it is equal on the real axis to the function $u(z,t)$ analytic in the lower half-plane. This analyticity requirement enforces chiral dynamics and hence results in chiral Burgers equation. We discuss this equation in more detail in Appendix~\ref{app:Burgers}. Here we just give a few examples of the solutions of (\ref{eq:almostBO-duplicate}).
%%%%%%%
\begin{figure}
\centering
\includegraphics[scale=0.6]{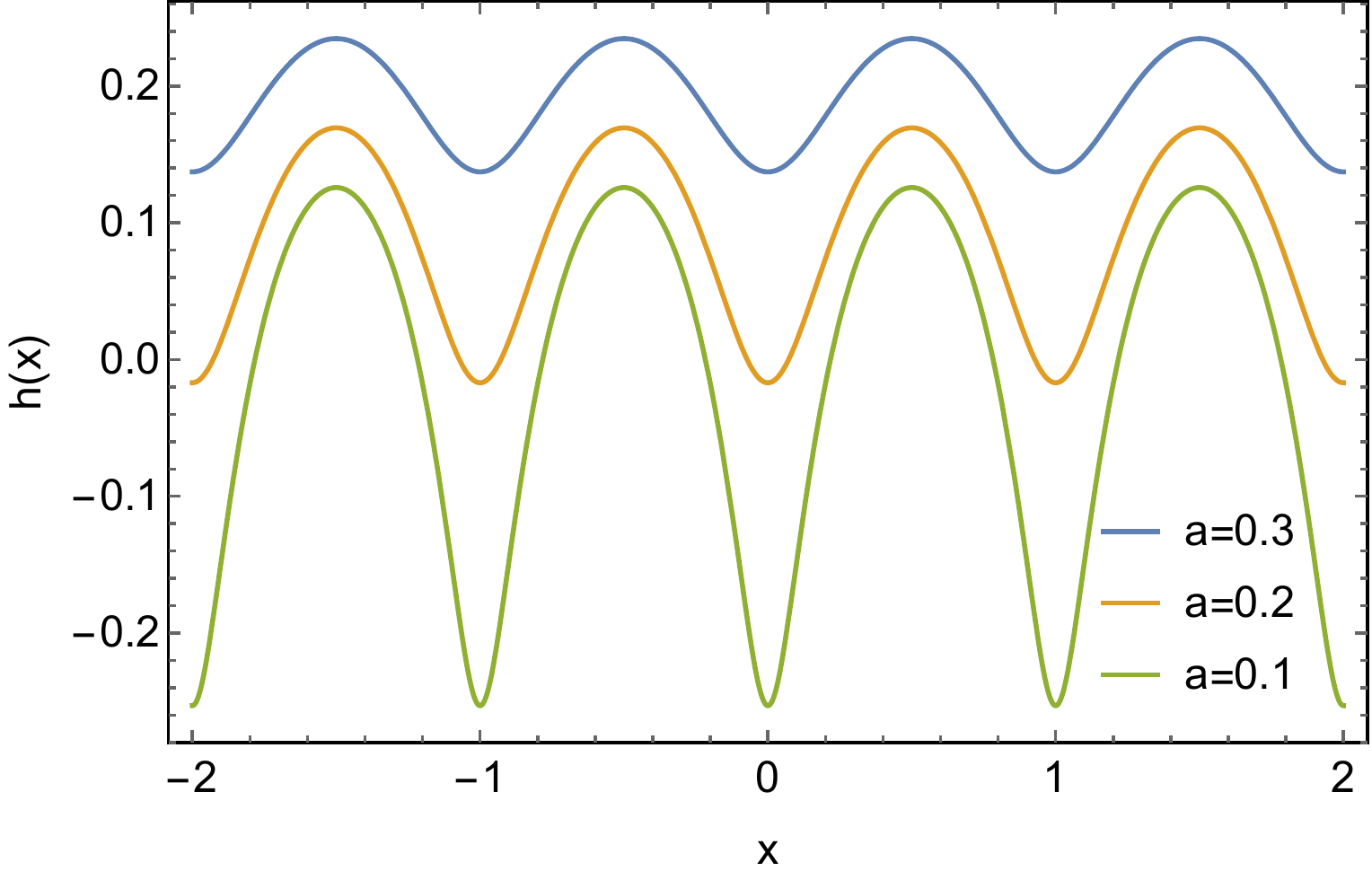}\\
\caption{\label{nlwave} Nonlinear surface wave profiles $h(x)$ given by Eq.~\ref{eq:h} for $k=2\pi$ and $a=0.1,0.2,0.3$ from bottom to top.}
\end{figure}
%%%%%%%
Let us start with the periodic nonlinear wave solution
\begin{align}
	u=\frac{U}{1-e^{i k(x-U t)+k a}}\,,	
 \label{eq:uminus}
\end{align}
where the velocity of the wave is given by 
\bea
	U=-2\nu_o k
\eea
and we restrict $k>0$ which guarantees the analyticity of (\ref{eq:uminus}) in the lower half of complex plane. The solution is characterized by two parameters: the wave vector $k>0$ and the parameter $a>0$. 

The real field $\tilde{u}=u+\bar{u}$ is then given by,
\begin{align}
	\tilde{u}=U\left(1-\frac{\sinh( k a)}{\cosh(k a)-\cos(k(x-Ut))}\right) \,.
 \label{eq:utilde}
\end{align}
We can also find the profile of the wave $h(x,t)$ from $\p_t h = \p_x\tilde\phi = i(u-\bar{u})$. We obtain integrating (\ref{eq:uminus}) over time (up to an additive constant)
\begin{align}
	h= \frac{1}{k}\ln\Big[\cosh(ka)-\cos(k(x-Ut))\Big] \,.
 \label{eq:h} 
\end{align}

%%%%%%%
\begin{figure}
\centering
\includegraphics[scale=0.6]{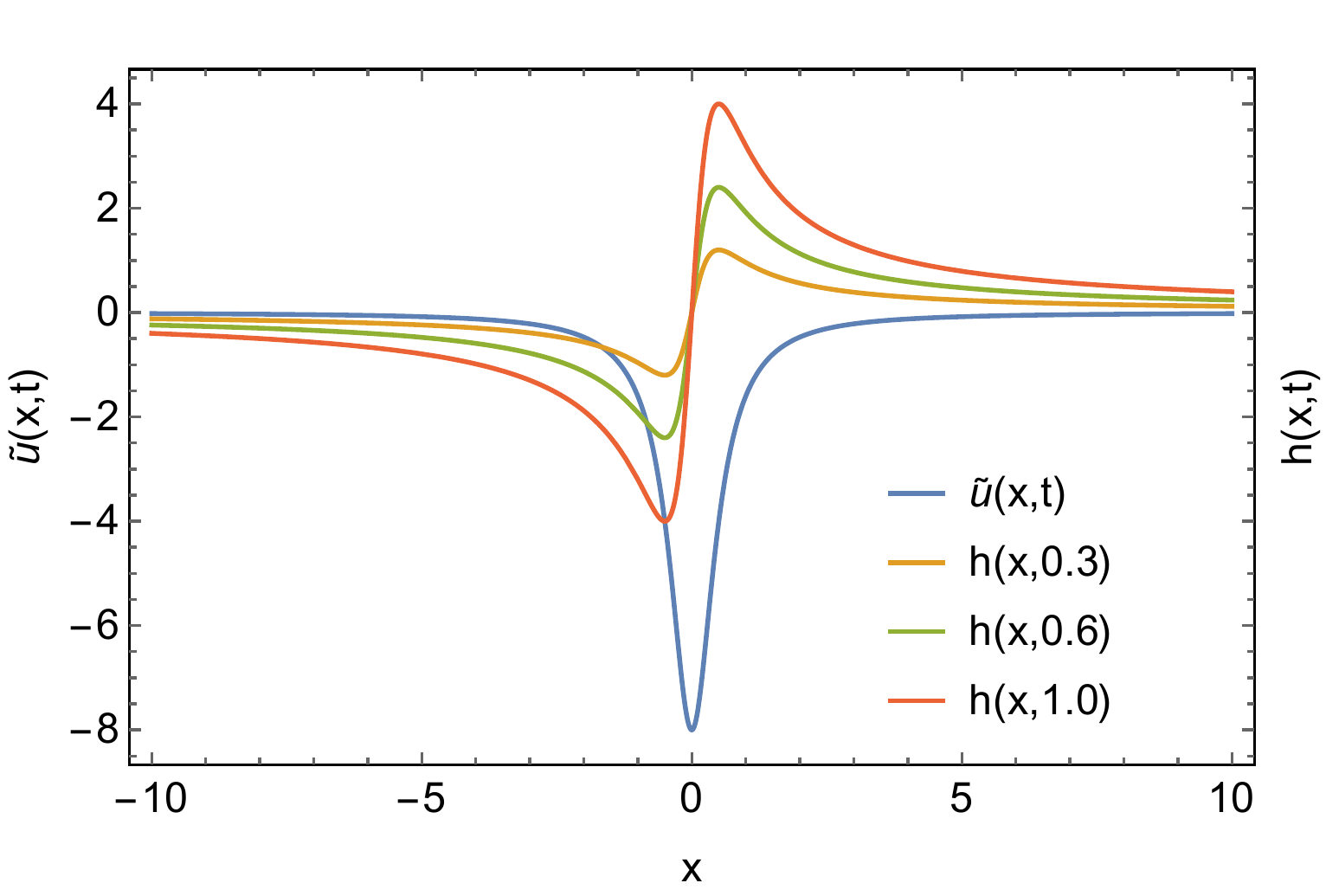}\\
\caption{\label{wavesingle} Nonlinear surface wave profiles $\tilde u(x),h(x,t)$ for different times $t=0.3,0.6,1.0$.}
\end{figure}
\begin{figure}
\centering
\includegraphics[scale=0.51]{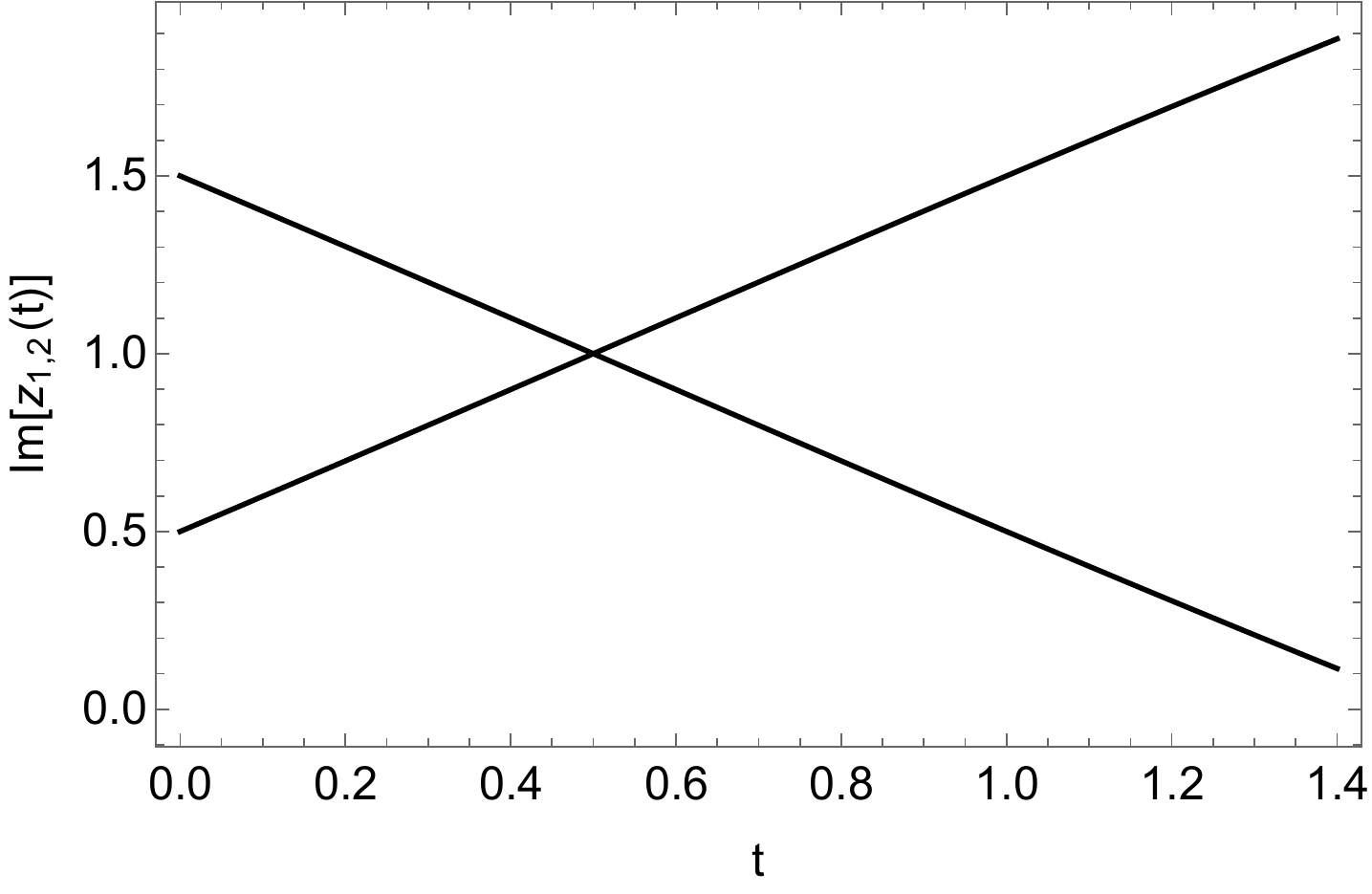}\quad \includegraphics[scale=0.51]{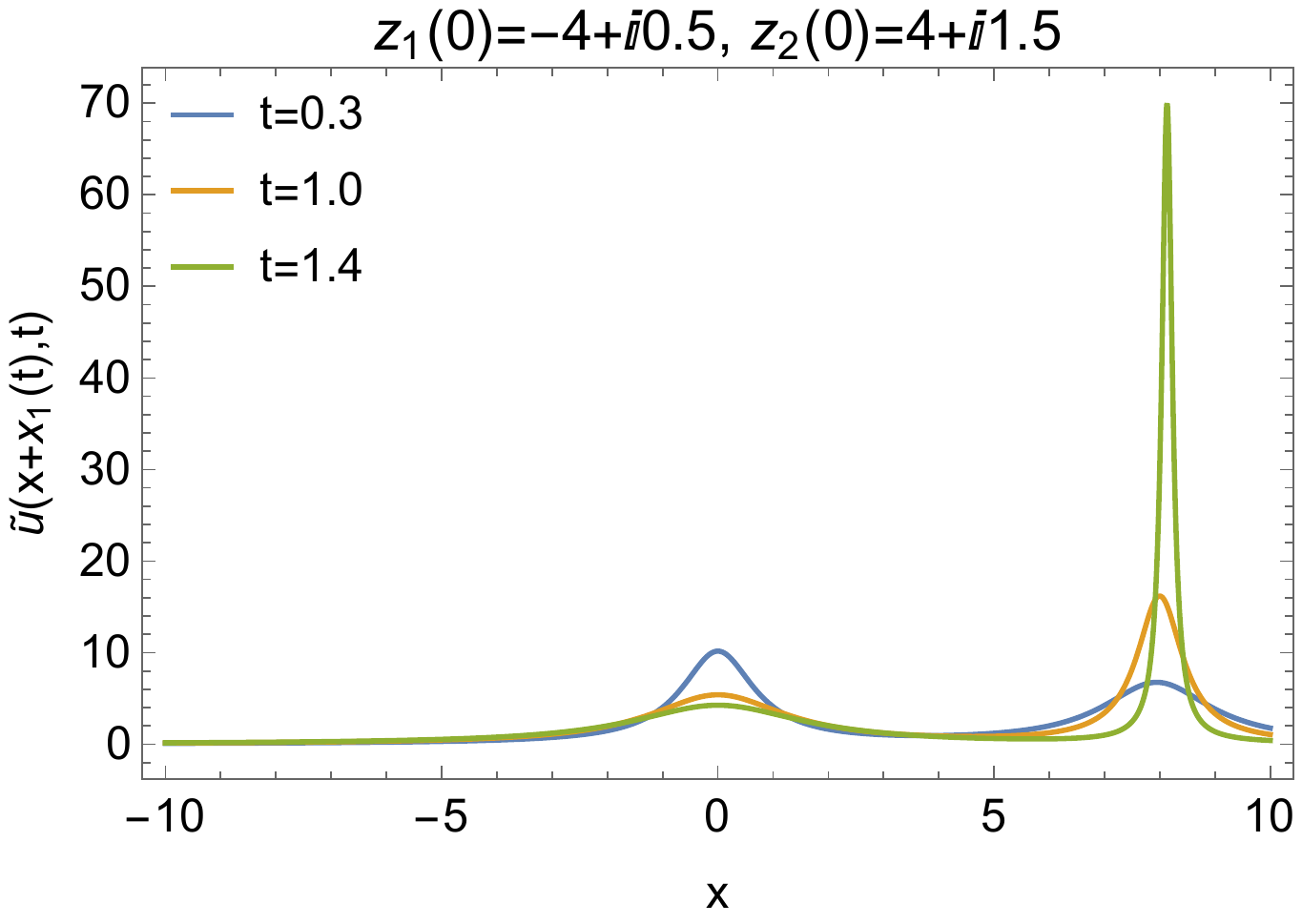}
\caption{\label{fig:2soliton} (left) Dynamics of the imaginary part of the two poles given by Eq.~\ref{eq:2soldyn}. One of the poles hits the real axis at finite time. (Right) Two soliton profile $\tilde u(x,t)$ for different times $t=0.3,1.0,1.4$ governed by the pole dynamics plotted on the left. One of the soliton profiles widens whose pole moves away from the real axis and the other soliton keeps sharpening while moving to the right (relative to the first soliton) and diverges at the time when the pole hits the real axis.}
\end{figure}

%%%%%%%
In the limit of $a\rightarrow \infty$ while keeping $k>0$ fixed, we obtain,
\begin{align}
	\tilde{u} \approx D\Omega \cos(kx-\Omega t) \,,\quad
 	h \approx D \cos(kx-\Omega t).
\end{align}
Here the amplitude of the wave $D=-2k^{-1}e^{-ka}$ and $\Omega=-2\nu_o k^2$. In this limit we recover the linearized Lamb solution of Section~\ref{sec:swodd}. Fixing $a$ and taking a limit $k\to 0$ we obtain the time independent solution for $u$ as
\begin{align}
	u = -\frac{2i\nu_o}{x+ia} \,,\quad \tilde{u} = -\frac{4\nu_o a}{x^2+a^2}\,.
\end{align}
This solution is reminiscent of the well known single-soliton solutions of the BDO equation7. However, in this limit we obtain the following expression for the height profile (see Fig.~\ref{wavesingle})
\begin{align}
	h &= \frac{4\nu_o x}{x^2+a^2} t \,.
\end{align}
This solution grows in time linearly. In time of the order of $t_* =a^2/\nu_o$ the curvature of the profile becomes significant and the assumptions of small curvature used in deriving (\ref{eq:almostBO-duplicate}) are violated. It is known that complex Burgers equation without the analyticity requirement (\ref{eq:almostBO-duplicate}) possesses multi-pole solutions \cite{senouf1996pole}. In Appendix~\ref{app:multipole} we present some of multi-pole solutions. Here we discuss just a two-pole solution. It is given by  
\begin{align}
	 u(x,t) = -2i\nu_o \left(\frac{1}{x-z_1(t)}+\frac{1}{x-z_2(t)}\right),
\end{align}
where
\begin{align}
	\dot z_{1,2}(t) = \mp\frac{ 4i\nu_o}{z_1-z_2} \,.
 \label{eq:2soldyn}
\end{align}
This solution is plotted in Figure~\ref{fig:2soliton} for several times. It essentially looks like two lumps which move and change their width. One of them is spreading while the other one is shrinking. At some finite time one of the complex poles hits the real axis and one cannot assume analyticity of $u(x)$ in the lower half complex plane beyond that time. 
%%%%%%%
%%%%%%%

%%%%%%%%%%%%%%%%%%%%%%%%%%%%
\section{Discussion and Outlook} 
%%%%%%%%%%%%%%%%%%%%%%%%%%%%

In this work we considered the motion of a free surface of an incompressible fluid with odd viscosity. For such an incompressible fluid the effect of the odd viscosity reduces to a modification of boundary conditions at the surface. 

We solved the full system of hydrodynamic equations with proper boundary conditions in the linear approximation in Section~\ref{sec:linoddwaves}. This solution generalizes the well known Lamb's result for linear surface gravity waves with ordinary shear viscosity~\cite{lamb1932hydrodynamics}. Similar to the Lamb's solution, the motion of the fluid is mostly potential in the bulk with a narrow vorticity layer formed at the surface due to the tangent stress present in viscous fluids. The structure of the  boundary layer is, however, different due to the presence of odd viscosity. The width of the boundary layer is $\sim \sqrt{\nu_e}$ as in Lamb's case, but vorticity in the layer diverges as $1/\sqrt{\nu_e}$ in the limit of vanishing shear viscosity $\nu_e\to 0$. The latter is in contrast with conventional Lamb's result where vorticity remains finite even within the boundary layer. Unlike the shear viscosity, the odd viscosity is non-dissipative and results in a real correction to the dispersion of surface waves. We computed these corrections in different limits. Remarkably, at finite odd viscosity the surface waves can propagate even in the absence of gravity with the dispersion of linear waves given by Eq.~\ref{eq:OmegaDisp}. The corresponding wave is chiral with the direction of the propagation defined by the sign of the odd viscosity coefficient $\nu_o$. In this paper we focused on the case when gravity is absent. 

The motion of the fluid corresponding to surface waves is potential except for the narrow boundary layer at the surface. One might try to represent the effect of such a boundary layer as modified boundary conditions imposed on irrotational (potential) fluid. We derived such boundary conditions and obtained a system of nonlinear Hamiltonian equations governing the dynamics of free surface. Remarkably, in the limit of vanishing shear viscosity, the only effect of the odd viscosity is a shift in the Poisson brackets (Eqs.~\ref{eq:PBphih} \ref{eq:PBphiphi} and \ref{eq:PBhh}) without any changes to the Hamiltonian (compare to Ref.~\cite{wiegmann2014anomalous}). In total we obtain three conservation laws corresponding to energy, mass and (pseudo)momentum. The dynamics corresponding to the Hamiltonian within small surface angle approximation reduces to  the chiral Burgers equation with imaginary viscosity (\ref{eq:almostBO}) as the effective one-dimensional equation governing weakly nonlinear surface waves in Section~\ref{sec:effheur} and Appendix~\ref{app:QPA}.  The chiral Burgers term results in non-singular solution in the form of periodic non-linear waves in contrast to the inviscid limit where almost all initial condition results in finite time singularities~\cite{kuznetsov1994formation}. However, even for chiral Burgers case generic pole solutions result in finite time singularities (see Appendix~\ref{app:multipole}). It would be interesting to study if and how the full non-linear Hamiltonian system alters these small angle finite time singularities. We leave this interesting study for future.

The complex Burgers equation has attracted some attention from a purely mathematical point of view as a simple example of a nonlinear dispersive equation \cite{dobrokhotov1992problem,senouf1996pole}. However, to the best of our knowledge the chiral Burgers dynamics of surface waves caused by odd viscosity is the first physical application of such an equation.

For vanishing shear viscosity, the vortical boundary layer becomes infinitesimally thin. With divergent vorticity, it is essentially equivalent to the discontinuity of the tangent velocity at the boundary. This means that the solution of the hydrodynamic equations for $\nu_e=0$ is a weak solution with a discontinuity of the tangent component of fluid velocity at the boundary. The vorticity is generated through odd viscosity by the motion of the boundary of the fluid.\footnote{This is similar to the known fact that in compressible fluids the vorticity is generated by compression through the odd viscosity \cite{avron1998odd}. In the incompressible case, the motion of the boundary plays the role of compression generating vorticity.} When shear viscosity is finite but very small, its only effect is to spread the vorticity layer to make its width $\sim \sqrt{\nu_e}$. In fact, instead of considering finite shear viscosity to regularize the weak solution one might put it to zero and instead consider the fluid with small but finite compressibility. In this case, the incompressible limit corresponds to the limit of infinite sound velocity $v_s\to \infty$ and the width of the boundary layer scales roughly as $\sim 1/v_s$. However, the structure of the boundary layer is different in this case. While the dispersion of linear surface waves in the limit $v_s\to \infty$ remains (\ref{eq:OmegaDisp}), finding the effective nonlinear equation is still an open question. We present an analysis of weakly compressible case  elsewhere.%\footnote{{\red We note here that as the odd viscosity is non-dissipative for the compressible fluid in the absence of dissipative viscosity a hydrodynamic variational principle can be used facilitating computations. We present details of the variational principle with odd viscosity elsewhere. This shows that the complex Burgers equation derived here is a universal equation for odd viscosity driven surface waves in the weakly nonlinear limit.[What is the purpose of this footnote exactly?]}}

It is known that the complex Burgers equation (\ref{eq:almostBO}) possesses multi-pole solutions~\cite{senouf1996pole}.  Not all of these solutions are acceptable for the chiral Burgers dynamics of  surface waves considered here. In fact, the chiral Burgers equation with additional analytic requirements comes from the real integro-differential equation (\ref{eq:almostBO1}). This equation  belongs to a one-parameter family of integro-differential equations together with the Benjamin-Davis-Ono equation  (see Appendix~\ref{app:family}). It deserves to be studied separately in more detail.

We consider some of the exact solutions of (\ref{eq:almostBO}) appropriate to our boundary conditions in Section~\ref{sec:nln} and Appendices~\ref{app:Burgers},\ref{app:multipole}. We find that some solutions, such as the periodic solution (\ref{eq:uminus},\ref{eq:utilde},\ref{eq:h})  are well behaved. On the other hand the multipole solutions exist only for finite time of the order $t_*=a^2/\nu_0$, where $a$ is some typical length scale. Within this time, one of the complex poles characterizing the solution hits the real axis, which corresponds to one of the bumps becoming very nonlinear. Close to this time the main assumption of weak nonlinear corrections we made in this work are violated. We cannot trust the chiral Burgers equation about and beyond this point in time. The behavior of the surface wave at longer times remains an open problem. There are a few scenarios that seem to be possible. One is that the equation can be corrected by terms higher order in derivatives and nonlinearity and those corrections prevent solutions from getting sharp and highly nonlinear. Another is that beyond $t_*$ the assumption of the vorticity confined to a boundary layer breaks severely and the layer is destroyed. Then the full two-dimensional dynamics should be used. The final scenario seems to be the most attractive to us. In this scenario the vorticity stops to be confined to a boundary layer but in a nice way. Namely, the imaginary pole hitting the real axis (surface of the fluid) penetrates inside the fluid forming a real vortex of quantized vorticity proportional to $\nu_o$. One can check this conjecture by repeating our derivations not assuming the full potentiality of the fluid in the two-dimensional bulk but allowing point vortices in the bulk. We postpone this study to the future.

We would like to conclude with a remark that while the study of surface waves in incompressible fluid with odd viscosity was partially motivated by hydrodynamics of quantum Hall states, the hydrodynamic equations considered here are not directly relevant to quantum Hall physics. While quantum Hall states are thermodynamically incompressible, they correspond to a vanishing sound velocity limit due to the gap in the excitation spectrum (in contrast to infinite sound velocity limit considered here). Of course, in addition to this subtle difference, the external magnetic field should definitely be included into quantum Hall hydrodynamics. We believe that the hydrodynamics equations considered in \cite{abanov2013effective} might serve as a good starting point to investigate surface modes of quantum Hall states. More directly, the results of this work should be applicable to chiral active liquids which are expected to have non-vanishing odd viscosity~\cite{banerjee2017odd}. In those applications one should also consider various types of damping terms such as rotational viscosity as well as other parity violating terms.

%%%%%%%%%%%%%%%%%%%%%%%%%%%%
\paragraph{Acknowledgements} 
%%%%%%%%%%%%%%%%%%%%%%%%%%%%

We are grateful to Gustavo Monteiro, William Irvine, Paul Wiegmann, and  Aleksandr Bogatskiy for many fruitful discussions related to this project. AGA's work was supported by grants NSF DMR-1606591 and US DOE DESC-0017662.

%\bibliography{oddviscosity-bibliography.bib}

%%%%%%%%%%%%%%%%%%%%%%%%%%%%
%%%%%%%%%%%%%%%%%%%%%%%%%%%%
%%%%%%%%%%%%%%%%%%%%%%%%%%%%
%%%%%%%%%%%%%%%%%%%%%%%%%%%%
%%%%%%%%%%%%%%%%%%%%%%%%%%%%
%%%%%%%%%%%%%%%%%%%%%%%%%%%%
\newpage

\appendix

%%%%%%%%%%%%%%%%%%%%%%
\section{Dispersion relation for linear surface waves}
 \la{app:dispersion}
%%%%%%%%%%%%%%%%%%%%%%

In this appendix, we discuss in more detail the dispersion relation for surface waves in various limits. The starting point for this analysis is Eqs. (\ref{eq:AB1}) and (\ref{eq:AB2}) which combined with (\ref{eq:mkOmega}) yields the dispersion $\Omega(k)$. We note that the linear solution (\ref{eq:linvx}-\ref{eq:AB2}) has a symmetry $k\to -k, \Omega\to -\bar{\Omega}, m\to \bar{m}$ etc., where bar denotes complex conjugation. Because of this symmetry, to avoid double counting we will keep $k>0$ in this section.  Another symmetry which follows from the consistency condition is $k \to - k$, $\nu_{o} \to - \nu_{o}$, $\Omega \to \Omega$, $m \to m$. This indicates that we can also stick to positive $\nu_{o} = |\nu_{o}|$. Then, the consistency of (\ref{eq:AB1},\ref{eq:AB2}) gives an equation for the dispersion
\begin{align}
	&\frac{g k - \Omega^{2} - 2 \Omega k^2(\nu_{o} + i \nu_{e} )}
	{2k^2(\nu_{o} +i \nu_{e})} 
	= \frac{g k - 2 \Omega k( \nu_{o} k  + i \nu_{e}  m)}{\Omega + 2 k(\nu_{o}  m+  i \nu_{e} k)}\,,
 \\
	&m^{2} = k^{2} - \frac{i \Omega}{\nu_{e}}\,, \quad {\rm Re} (\tilde{m}) > 0 \,.
\end{align}
The first equation can be turned into a sixth order polynomial equation for 
 $m$ using the second equation.

One possible solution of this equation, which can be verified by direct inspection, is $m=k$ and $\Omega = 0$ -- a stationary mode. We can factorize the polynomial equation for $m$  to exclude this solution. The result is a fifth order equation. Introducing the dimensionless quantities

\begin{align}
	\beta^2 = \frac{\nu_{e} k^{2}}{\sqrt{gk}}\,,  \quad { \alpha = \frac{\nu_{o}}{\nu_{e}}}\,, \quad \kappa = m/ k, \quad \Omega = i \nu_{e} k^{2} (\kappa^{2} - 1),
\end{align}

we express the equation for the dispersion in the following economical form

\begin{align}
	\beta^{-4}  ( \kappa + 1 - 2 i \alpha) 
	+  (\kappa+1) \left[  \kappa^{2} + 1 - 2i\alpha  \right] \left[ \kappa^{2} 
	+1-    2i \alpha \kappa \right] 
	= 4 (\kappa + 1) (\kappa- i \alpha) (1 - i \alpha)\,.
 \label{eq:polydisp}
\end{align}

We will proceed now to solve (\ref{eq:polydisp})  perturbatively in various limits. 

%%%%%%%%%%%%%%%%%%%%%%
\subsection{Overdamped Waves} In the overdamped regime with $g = 0$ and $\nu_{o} = 0$, the problem reduces to
\begin{align}
	\kappa^{4} +2 \kappa^{2} + 1 -4 \kappa = 0\,,
\end{align}
which actually has $\kappa = 1$ as a solution. This means the stationary zero mode is doubly degenerate. There is an additional root with real positive part ${\rm Re} (\kappa) > 0$. This means there is a single overdamped mode with imaginary quadratic dispersion
 \begin{align}
 	\Omega &= -i \nu_{e} \left(1- \kappa_{0}^{2} \right) k^{2},
 \label{eq:overdamped}\\
 	\kappa_{0}  &= \frac{1}{3} \left( - 1 - \frac{4 \times 2^{2/3}}{\left( 13 + 3 \sqrt{33}\right)^{1/3}} 
	+ 2^{1/3}\left( 13 + 3 \sqrt{33}\right)^{1/3}\right)
	\approx 0.295598 \,.
 \end{align}
As anticipated, these waves are damped since ${\rm Im}(\Omega) <0$. This solution is perhaps remarkable for how complicated the numerical factor ends up being for the simplest case of an overdamped viscous wave. Note that $|1-\kappa_{0}^{2}|<1$.

\subsection{Gravity Dominated Surface Waves} Next, we consider a regime where $\beta << 1$, i.e. which is gravity dominated, and look at even and odd viscosity corrections to the classical gravity wave dispersion. 
	
	In order to fruitfully analyze the polynomial equation (\ref{eq:polydisp}), we will rescale $\kappa = \beta^{-1} x$ and consider small $\beta$ solutions of
	
\begin{align}
 ( x + \beta - 2 i \alpha \beta) +  ( x+\beta) \left[  x^{2} + \beta^{2} - 2i\alpha \beta^{2}  \right]\left[ x^{2} +\beta^{2}-    2i \alpha \beta x \right] = 4 \beta^{3} (x + \beta ) (x- i \alpha \beta) (1 - i \alpha)
 \la{eq:xbeta}
\end{align}

	\smallskip
 {\bf Zero Viscosity} To recover the classical dispersion, we simply set $\beta = 0$ in the equation above, yielding $x^{4} = -1$ and use $\Omega = \frac{i \nu_{e} k^{2}}{\beta^{2}} \left( x^{2} - \beta^{2}\right) \to 	i \sqrt{ g k} \,x^{2}$ to give dispersion for deep gravity waves
\begin{align}
	\Omega = \pm \sqrt{g k}\,, 
 \label{eq:gravitydisp}
\end{align}
In this limit, there is only a single vertical length scale, so $m$ completely drops out of the problem. 

\smallskip

{\bf Lamb's Solution} Keeping the odd viscosity zero, but turning on a small shear viscosity amounts to considering $\alpha = 0$ and small $\beta <<1$ corrections to the dispersion (\ref{eq:gravitydisp}). Apparently, $x = -\beta$ is a solution. But since this is unphysical (negative solutions do not decay into the bulk), we simply factor it out and are left with 

\begin{align}
1 + (x^{2} + \beta^{2})^{2} = 4 \beta^{3}x,
\end{align}
The decaying solutions have the perturbative expansion
\begin{align}
	x_{\pm} = e^{\pm i \pi / 4}  \pm  \frac{i\beta^{2}}{2} e^{i \pi  / 4} + O(\beta^{3})\,,
\end{align}
which leads to the dispersion 
\begin{align}
	\Omega_{\pm}  &= \mp \sqrt{g k} - 2 i \nu_{e} k^{2} + O(\nu_{e}^{3/2}),
\end{align}
Curiously, damped gravity waves decay faster than the overdamped mode (\ref{eq:overdamped}).

\smallskip

{\bf Odd Corrections to Lamb's solution} We consider corrections to Lamb's solution when odd viscosity is the smallest scale, i.e. $\alpha << \beta <<1$. In this case, there are again two propagating modes 
\begin{align}
	x_{1} &=  - e^{ 3i \pi/4} + \frac{1}{2}\beta^{2} e^{ 3 i \pi /4} (i + 2 \alpha) -  i\beta ^{3}  (i + \alpha)^{2} + O(\beta^{4}) \,,\\
	x_{2} & = e^{ i \pi /4} + \frac{1}{2}\beta^{2}  e^{ i \pi /4} ( i + 2 \alpha) +  i\beta^{3} (i + \alpha)^{2} + O(\beta^{4}) \,.
\end{align}

%	\begin{align}
%	x_{1}^{2} & = e^{3 i \pi / 2} - 2 \beta^{2} e^{ 3 i \pi /2} ( i + 2\alpha) + 12 i \beta^{3} e^{ 3i \pi /4} ( i + \alpha)^{2} + ...	\\
%	x_{2}^{2} & = e^{ i \pi / 2} + 2 \beta^{2} e^{  i \pi /2} ( i + 2\alpha) + 12 i \beta^{3} e^{ i \pi /4} ( i + \alpha)^{2} + ...	
%	\end{align}

\begin{align}
	\Omega_{1} & %=i \sqrt{g |k|} \left( - i +i 2 \beta^{2} 	 ( i + 2\alpha) - \beta^{2}\right) = \sqrt{ g |k|} - \nu_{e} k^{2} (3i + 4 \frac{\nu_{o}}{\nu_{e}} {\rm sign}( k) 
	=\sqrt{g k} - 2 i \nu_{e} k^{2} - 2 \nu_{o} k^2 +O(\nu_{e}^{3/2}) \,,
 \la{eq:LOmega1} \\
	\Omega_{2} & = - \sqrt{g k} - 2 i \nu_{e} k^{2} - 2 \nu_{o} k^2 + ...\,.
 \la{eq:LOmega2}
\end{align}

This is technically a perturbation series in $\beta$ having fixed $\alpha$. Therefore, it is valid also in the regime where $\nu_{e} k^{2}\ll \nu_{o} k^2\ll \sqrt{g k}$. 
	
\smallskip

{\bf Zero Shear Viscosity} Finally, we find the dispersion in the absence of shear viscosity but without any additional assumptions on gravity and odd viscosity. We take a limit $\alpha\to \infty$, $\beta\to 0$ keeping 
\begin{align}
	\mu = \alpha\beta^2= { \frac{\nu_o k^2}{\sqrt{gk}}} = const\,.
\end{align}
We obtain from (\ref{eq:xbeta}) the equation $x^4-2i\mu x^2+1=0$ and corresponding dispersion
\begin{align}
	\Omega =\sqrt{gk}\Big(\pm \sqrt{1+\mu^2}-\mu\Big)\,.
 \la{eq:Omegamu}
\end{align}
The latter formula produces (\ref{eq:gravitydisp}) for $\mu=0$ ($\nu_o=0$) and gives corrections to Lamb's solution (\ref{eq:LOmega1},\ref{eq:LOmega2}) for $\nu_e=0$. However, Eq.~\ref{eq:Omegamu} can also be used when gravity is negligible producing in this limit $\Omega=\{0,-2\nu_o k^2\}$ (c.f., with the next section).

%%%%%%%%%%%%%%%%%%%%%%
\subsection{Viscosity Dominated Surface Waves} The main case considered in the bulk of the paper consists of viscosity dominated waves with  $\nu_e\ll |\nu_o|$, and  $g=0$. For zero gravity, the equation (\ref{eq:polydisp}) simplifies considerably. In fact, using $\Omega = \tilde{\Omega} k^{2}$, we find that $\tilde{\Omega}$ is independent of $k$ and given by
	
\begin{align}
	&\frac{- \tilde{\Omega}^{2} - 2 \tilde{\Omega}( \tilde{\nu}_{o}  + i \nu_{e} )}{2(\tilde{\nu}_{o}  + i \nu_{e})} = \frac{- 2 \tilde{\Omega}  ( \tilde{\nu}_{o}  + i \nu_{e}  \tilde{m})}{\tilde{\Omega} + 2 ( \tilde{\nu}_{o}  \tilde{m}+  i \nu_{e} )}, \quad \tilde{m} = \sqrt{1 - i \tilde{\Omega}/\nu_{e}}
\end{align}
We find that the leading dispersion in the limit $\nu_{e}\ll|\nu_{o}|$ is given by

\begin{align}
	\Omega %&= - 2 a q^{2}  - i r q^{2} \left(  1 + i\chi \right) \frac{1}{b} - \frac{9}{4} i r q^{2}\\
	& \approx 	- 2 \nu_{o} k^2 - (i - {\rm sign}(\nu_{o} ) )  k^{2} \sqrt{|\nu_{o}| \nu_{e}} 
	- \frac{3}{2} i \nu_{e} k^{2} + O(\nu_{e}^{3/2}) 
 \nonumber \\
 	&\approx - 2 \nu_{o} k^2 - i k^{2} \sqrt{|\nu_{o}| \nu_{e}}\,,
\end{align}
which describes a chiral wave propagating in the direction set by the sign of $\nu_{o}$. The boundary layer is characterized by  
\begin{align}
	m&=  k\sqrt{ \frac{|\nu_{o}|}{\nu_{e}}}  
	\left( 1 + i\,{\rm sign}(\nu_{o} )   - \frac{1}{2} \sqrt{\frac{\nu_{e}}{|\nu_{o}|}} 
	+ \frac{3}{16} (  i\, {\rm sign}(\nu_{o} ) - 1) \frac{\nu_{e}}{|\nu_{o}|} +  O(\nu_{e}^{3/2})\right)
%m & = |k| \sqrt{ \frac{|\nu_{o}|}{\nu_{e}}} ( 1+ \chi i ) - \frac{1}{2}  |k|
\end{align}
showing that the width of the boundary layer for disturbances of characteristic length $\lambda$ scales like $\lambda \sqrt{\nu_{e}/ |\nu_{o}|}$. Furthermore, we see that the imaginary part will lead to a rapidly varying phase in the vertical direction. 

%%%%%%%%%%%%%%%%%%%%%%
\section{Harmonic functions and Hilbert transform}
 \la{app:hilbert}
%%%%%%%%%%%%%%%%%%%%%%

A function $\phi(z=x+iy)$ harmonic for $y\leq h(x)$ can be written as
\begin{align}
	\phi(z) = \R \frac{1}{\pi}\int_{-\infty}^{+\infty}dx' \rho(x') \log\Big(z-x'-ih(x')-i0\Big),
\end{align}
 where i0 is equivalent to $i\epsilon$ with $\epsilon\rightarrow +0$. We calculate
\begin{align}
	(\p_y\phi)_{x+ih} &= \R \frac{i}{\pi}\int_{-\infty}^{+\infty}dx' \rho(x') \frac{1}{x-x'+i(h(x)-h(x'))-i0}
 \nonumber \\
    & \approx -\rho+h\rho_x^H-(h\rho)_{x}^{H} +O(h^2) \,,
\end{align}
where we have used the definition of Hilbert transform given in Eq.~\ref{eq:Htransf} and we used Plemelj formula,
\begin{align}
\frac{1}{x-x'-i0}=P.V.\frac{1}{x-x'}+i\pi \delta(x-x')\nonumber
\end{align}
such that,
\begin{align}
\frac{1}{\pi}\int dx'\frac{\rho(x')}{x-x'-i0}=-\rho^H(x)+i\rho(x).	\nonumber
\end{align}
Then we introduce $\tilde\phi(x) =\phi(x,h(x))$ and compute
\begin{align}
	\p_x\tilde\phi = \R \frac{1}{\pi}\int_{-\infty}^{+\infty}dx' \rho(x') \frac{1+ih_x(x)}{x-x'+i(h(x)-h(x'))-i0}\approx -\rho^H+O(h^2) \,.
\end{align}
Comparing these two equations we obtain to linear order in $h$
\begin{align}
	(\p_y\phi)_{x+ih}=-\tilde\phi_x^H - h\tilde\phi_{xx} -(h\tilde\phi_x^H)_x^H  \,.
 \la{eq:phiy}
\end{align}
This form coincides with Kuznetsov et al. \cite{kuznetsov1994formation}. 
For completeness we also give
\begin{align}
	(\p_x\phi)_{x+ih}=\tilde\phi_x +h_x\tilde\phi^H_x \,
 \la{eq:phiy1}
\end{align}
and another useful formula valid to linear order in $h$
\begin{align}
	(\p_y\phi-h_x\p_x\phi)_{x+ih}=-\tilde\phi_x^H - (h\tilde\phi_{x})_x -(h\tilde\phi_x^H)_x^H \,.
 \la{eq:phiycomb}
\end{align}

%
%Kinematic boundary condition becomes
%\begin{align}
%	h_t +\tilde\phi_x h_x \approx -\tilde\phi_x^H - h\tilde\phi_{xx} -(h\tilde\phi_x^H)_x^H
%\end{align}
%or 
%\begin{align}
%	h_t +\tilde\phi_x^H+\Big[\tilde\phi_x h +(h\tilde\phi_x^H)^H\Big]_x \approx 0 \,.
%\end{align}
%In terms of $\tilde{u}=\tilde\phi_x$ it becomes
%\begin{align}
%	h_t +\tilde{u}^H+\Big[\tilde{u} h +(\tilde{u}^H h)^H\Big]_x \approx 0 \,.
%\end{align}

%%%%%%%%%%%%%%%%%%%%%%%%%%%%%
\subsection{Useful formulas}
%%%%%%%%%%%%%%%%%%%%%%%%%%%%%

Using the decomposition $f^H=i(f^+-f^-)$, where $f^{\pm}$ is the function analytic in the upper(lower) half of complex plane one can derive the following identity:
\begin{align}
	(fg)^H-(f^Hg^H)^H = f^Hg+f g^H\,.
 \la{eq:fgH}
\end{align}
Another useful formula
\begin{align}
	f^H g^H-(f^Hg)^H = (f^+g^-)^+ +(f^-g^+)^-\,.
\end{align}
We conclude this section by noticing that the integral of total Hilbert transform vanishes
\begin{align}
	\int_{-\infty}^{+\infty} f^H(x) \,dx =0\,.
\end{align}
Applying this to (\ref{eq:fgH}) we can ``integrate by parts'' expressions with Hilbert transform
\begin{align}
	\int_{-\infty}^{+\infty} f(x) g^H(x) \,dx = -\int_{-\infty}^{+\infty} f^H(x) g(x) \,dx\,.
\end{align}
%%%%%%%%%%%%%%%%%%%%%%
\section{Quasi-potential approximation for nonlinear surface waves}
 \la{app:QPA}
%%%%%%%%%%%%%%%%%%%%%%

The goal of this section is to derive an effective nonlinear description of surface waves in the absence of gravity and shear viscosity. The main idea is to ``integrate out'' the vortical component of the fluid leaving equations that contain only the potential part of the flow -- ``quasi-potential approximation''. Technically we follow the works \cite{ruvinsky1991numerical,dias2008theory}. However, our results are very different from those works as the main physics is governed by the odd viscosity and we take a limit of vanishing shear viscosity.

%%%%%%%%%%%%%%%%%%%%%%
\subsection{Navier-Stokes equation}

In the following it is convenient to separate the velocity of the fluid into potential and vortical part
\be
	v_i = \p_i\phi +v^\psi_i = \p_i\phi +\p_i^*\psi\,,
\ee
where $\phi$ is the velocity potential and $\psi$ is a stream function. The velocity potential is harmonic $\Delta\phi=0$ as a consequence of incompressibility of the fluid.
We rewrite the Navier-Stokes equation (\ref{eq:euler10}) as
\be
	\p_t(\p_i\phi+v^\psi_i) -v_i^*\omega
	=-\p_i\left(\tilde{p}+\frac{v_j^2}{2}\right) +\nu_e \Delta v^\psi_i \,.
\ee
We integrate the $y$-component of this equation in $y$ from $-\infty$ to $h(x,t)$ and obtain
\be
	\p_t\phi + \frac{1}{2}(\vec\nabla\phi +{\vec v}_\psi)^2 +\tilde{p}+\int^{h}_{-\infty} v_x\omega dy
     = 
    -\int^{h}_{-\infty} (\p_tv_y^{\psi}-\nu_e \Delta v_y^{\psi})dy\,.
 \la{eq:Bintegr}
\ee
Let us now make the following assumptions in the limit $\nu_e\to 0$ (i) both potential and vortical components of velocity remain finite; (ii) the vortical component of velocity is essentially non-zero only in the narrow layer of thickness $\delta\sim \sqrt{\nu_e}$ near the boundary; (iii) the velocity of the fluid tangent to the boundary vanishes at the free surface; (iv) the vortical component of the velocity normal to the boundary vanishes.
 
The conditions (i) and (ii) mean that the vorticity might diverge as $1/\delta\sim \nu_e^{-1/2}$. Nevertheless, it is clear that the right hand side of (\ref{eq:Bintegr}) vanishes in the limit of $\nu_e\to 0$. However the term $\int^{h}_{-\infty} v_x\omega dy\approx\int^{h}_{-\infty} v_x\p_xv_y dy$ does not vanish in this limit in the presence of odd viscosity and results in a non trivial contribution in the form of,
\begin{align}
\int^{h}_{-\infty} v_x\p_xv_y\,dy
    &=\int^{h}_{-\infty} (v^\psi_x+\p_x\phi)\p_x(\p_y\phi+ v^\psi_y)\,dy
    \approx\int^{h}_{-\infty} (v^\psi_x+\p_x\phi)(\p_x\p_y\phi)\,dy
 \nonumber \\
    &\approx \int^{h}_{-\infty} (\p_x\phi)(\p_x\p_y\phi)\,dy
    =\frac{1}{2}(\p_x\phi)^2\bigg|_{y=h}.
\end{align}

Using the conditions (iii) and (iv) in the left hand side of (\ref{eq:Bintegr}) we obtain the following approximate condition at the boundary $y=h(x,t)$
\be
	\p_t\phi +\frac{1}{2}(\p_x\phi)^2+\frac{1}{2}(\p_y\phi)^2 +\tilde{p} = 0 \,.
 \la{eq:Bbound}
\ee
In the remainder of this section we will derive an expression for $\tilde{p}$ in terms of the potential $\phi$ and justify some of our approximations.

Let us introduce a local coordinate system $(s,n)$ as shown in Fig.~\ref{fig:sn}. In these curvilinear coordinates the tangential and normal boundary conditions are given by,
\begin{align} 
	T_{nn} &= -p +2\nu_e \p_n v_n +\nu_o (\p_s v_n+\p_nv_s-\kappa v_s) = 0\,,
 \la{eq:Tnn} \\
 	T_{sn} &= \nu_e(\p_s v_n+\p_nv_s-\kappa v_s) -2\nu_o \p_nv_n= 0 \,,
 \la{eq:Tsn} 
\end{align}
where expressions for vorticity and incompressibility conditions are given by
\begin{align} 
	\omega &=\partial_nv_s-\partial_sv_n+\kappa v_s \,,
 \la{eq:omegasn} \\
 	\partial_i v_i &= \p_n v_n+\p_s v_s +\kappa v_n = 0 \,.
 \la{eq:incomsn}
\end{align}

%%%%%%%%%%%%%%%%%%%%%%
\subsection{Tangent stress boundary condition}

Using (\ref{eq:omegasn},\ref{eq:incomsn}) we rewrite the tangent stress boundary condition (\ref{eq:Tsn}) as
\begin{align} 
	\p_sv_s+\kappa v_n &= -\frac{\nu_e}{\nu_o}\left(\p_sv_n-\kappa v_s+\frac{1}{2}\omega\right) \,,
 \la{eq:omegaTsn} 
\end{align}
Taking a limit $\nu_e\to 0$ and assuming that the divergence of $\omega \sim \nu_e^{-1/2}$ we obtain
\begin{align} 
	\p_sv_s+\kappa v_n &= O(\sqrt{\nu_e})\to 0 \,.
 \la{eq:vsboundary} 
\end{align}
This condition is consistent with the previously used assumption that $v_s=0$ in linear approximation when $\nu_e\to 0$. We will use the fact that $v_s$ is of at least the second order in the amplitude of the wave. 

%We also know that $v_n$ can be of the order of one and does not go to zero with even viscosity. This means that the curvature of the droplet boundary should scale as $\sqrt{\nu_e}$ as well. Let us know assume this scaling and come back to exact relation (\ref{eq:omegaTsn}) and drop all terms of the order higher than $\sqrt{\nu_e}$. We obtain:
%\begin{align} 
%	\p_sv_s+\kappa v_n &= -\frac{\nu_e}{2\nu_o}\omega \,.
% \la{eq:vsboundary2} 
%\end{align}
%We can think of this boundary condition as the one determining the boundary value of vorticity $\omega$.

%%%%%%%%%%%%%%%%%%%%%%
\subsection{Kinematic boundary condition}

The kinematic boundary condition (\ref{eq:KBC}) can be rewritten as 
\bea
	h_t  = v_n = \p_n\phi +v_n^\psi\,.
 \la{eq:KBC2}
\eea
Neglecting $v^\psi_n$ and using $\p_n \phi\approx \p_y \phi-h_x\p_x \phi$ we obtain:
\bea
	h_t = \p_y\phi-h_x\p_x\phi
 \la{eq:KBC3}
\eea
as a kinematic boundary condition. 

%%%%%%%%%%%%%%%%%%%%%%
\subsection{Normal stress boundary condition}

Let us now rewrite the boundary condition for the normal part of the stress tensor as
\begin{align} 
	T_{nn} &= -\tilde{p} +2\nu_o(\p_sv_n-\kappa v_s) +2\nu_e \p_n v_n  = 0 \,.
\end{align}	
Here we introduced $\tilde{p}=p-\nu_o\omega$ and used (\ref{eq:omegasn}). We immediately obtain for the boundary value of the modified pressure
\begin{align}
 	\tilde p& =2\nu_o(\p_s v_n-\kappa v_s)+2\nu_e\p_n v_n.
\end{align}
 Taking a limit $\nu_e\rightarrow 0$ and ignoring $\kappa v_s$ term\footnote{The tangent velocity in the limit $\nu_e\to 0$ is of the second order of smallness in the amplitude of the wave.} the higher order terms in this expression we obtain 
\begin{align}
	\tilde p& \approx 2\nu_o \p_s v_n 
\end{align}
and we can use it in (\ref{eq:Bbound}) to obtain
\be
	\p_t\phi + \frac{1}{2}(\p_x\phi)^2+\frac{1}{2}(\p_y\phi)^2 +2\nu_o\p_s v_n = 0 \,.
 \la{eq:Bboundphi10}
\ee
We proceed neglecting all terms smaller than quadratic ones as $\p_s v_n = (-\p_x-h_x\p_y) v_n = -\p_xv_n -h_x\p_y v_y=-\p_x v_n+h_x\p_x v_x = -\p_x v_n = -\p_x (\p_y\phi-h_x\p_x\phi)$. Here we again used the fact that $v_s$ and $v_x$ are quadratic in the amplitude. Using this relation we can transform (\ref{eq:Bboundphi10}) to 
\be
	\p_t\phi + \frac{1}{2}(\p_x\phi)^2+\frac{1}{2}(\p_y\phi)^2 
	=2\nu_o\p_x (\p_y\phi-h_x\p_x\phi)  \,.
 \la{eq:Bboundphi}
\ee
This is the equation that we derived in Sec.~\ref{sec:effheur} Eq.~\ref{eq:Bernoulli4}. This equation can be expressed in terms of  $\tilde \phi(x,t)=\phi(x,h(x,t),t)$ leading to the non-linear Hamiltonian system defined in Eqs.~\ref{eq:Bernoulli5} and \ref{eq:ht2}.

%%%%%%%%%%%%%%%%%%%%%%

\begin{comment}
\subsection{Effective nonlinear boundary dynamics}

Let us now introduce the boundary value of the velocity potential as
\bea
	\tilde\phi (x,t) = \phi(x,h(x,t),t) \,.
\eea
We would like to replace $\phi$ in (\ref{eq:Bboundphi}) by $\tilde\phi$. First of all, the function $\phi(x,y)$ is harmonic for $y\leq h(x,t)$  and, therefore we can use (\ref{eq:phiy}) in linear terms and $\p_y\phi\approx -\tilde\phi^H_x$ in all quadratic terms.
For the time derivative we have
\bea
	(\p_t \tilde\phi)_x = (\p_t \phi(x,h(x,t),t))_x = (\p_t \phi(x,y,t))_{x,y} +(\p_y\phi)h_t
	\approx (\p_t \phi(x,y,t))_{x,y} +(\p_y\phi)^2 \,.
 \la{eq:ptphi}
\eea
Here subscripts show which variables are kept constant at taking partial derivatives. We substitute (\ref{eq:phiy},\ref{eq:ptphi}) into (\ref{eq:Bboundphi}) and obtain
\be
	\tilde\phi_t+\frac{1}{2}(\tilde \phi_x^2 -(\tilde\phi_x^H)^2)+2\nu_o\tilde\phi^H_{xx} 
	= -2\nu_o\Big[h\tilde\phi_{x}+(h\tilde\phi_x^H)^H\Big]_{xx}    \,.
 \la{eq:eff}
\ee
This equation is the main result of this section.
We also use (\ref{eq:phiy},\ref{eq:ptphi}) in (\ref{eq:KBC3}) to obtain the kinematic boundary condition in the form
\be
	h_t +\tilde\phi_x^H=-\Big[h\tilde\phi_x  +(h\tilde\phi_x^H)^H\Big]_x\,.
 \la{eq:ht10}
\ee
This equation coincides with the kinematic boundary condition in Ref.~\cite{kuznetsov1994formation}.

}
\end{comment}
%%%%%%%%%%%%%%%%%%%%%%
\section{Nonlinear periodic chiral surface waves in chiral Burgers equation}
 \la{app:Burgers}
%%%%%%%%%%%%%%%%%%%%%%
The goal of this appendix is to derive a periodic moving wave solution of the chiral Burgers equation  (\ref{eq:almostBO-duplicate}) used in Section~\ref{sec:nln}.
%
% 
%We now show that (\ref{eq:almostBO1}) is related to complex Burgers equation. We introduce $\phi(x)=\phi^+(x)+\phi^-(x)$, where $\phi^\pm(z)$ is the function analytic in upper (lower) half-planes of $z$. Then the Hilbert transform (\ref{eq:Htransf}) is equivalent to $\phi^H=i(\phi^+-\phi^-)$. In Fourier space $\phi^H(k) = i\sign(k) \phi(k)$. The function $\Phi(z)=\phi +i\phi^H$ is holomorphic in the lower half plane if $\phi(x)$ is harmonic in the lower half plane. Therefore, $\phi_y=-\phi_x^H$ which we used in deriving \ref{eq:Bernoulli3}. 
%
%This equation can be decoupled into $\phi^{\pm}$ as,
%
%\be
%	\phi^{\pm}_t+(\phi^{\pm}_x)^2 \pm 2i\nu_o\phi^{\pm}_{xx} = \pm iQ(x) \,.
% \la{eq:decoup}
%\ee
%Where $Q(x)$ is a generic analytic function. Consider the class of solutions of the form $\phi^-\equiv f(x-U t)$ which has poles only in the upper half plane. In order to satisfy the boundary conditions at $y=0$ and at $y=-\infty$, we require $Q(x)=0$. Rewriting the above equation in terms of the velocity field $u=\phi_x$, we obtain,
%\begin{align}
%	u^{\pm}_t+2u^{\pm} u^{\pm}_x \pm 2i\nu_o u^{\pm}_{xx} = 0 \,.
% \la{eq:decoupvel}
%\end{align}
%The above equation then takes the form of the complex Burgers equations with an additional reality condition that $u^{\pm}$ is analytic in the upper and lower half plane respectively such that $u^-+u^+=u$ is real. 
%
%
We look for a moving wave solution of the form $u=u(x-Ut)$. We substitute it into (\ref{eq:almostBO-duplicate}) and obtain:
\begin{align}
	\partial_x\Big(-U u+u^2-2i\nu_o u_{x}\Big)=0\,.
\end{align}
Integrating the above equation and setting the overall constant as $A$ we obtain,
\begin{align}
	-U u+u^2-2i\nu_o u_{x}=A
\end{align}
Integrating this equation gives:
%The above equation can be rewritten as,
%\begin{align}
%\frac{1}{2i\nu_o }=\frac{u^-_{x}}{(u^-)^2-U u^--A}=\frac{u_x^-}{U_1-U_2}\left(\frac{1}{u^--U_1}-\frac{1}{u^--U_2}\right)
%\end{align}
\begin{align}
	\frac{2i\nu_o}{U_1-U_2}\log\left(\frac{u-U_1}{u-U_2}\right)	=x-x_0-Ut-ia\,.
\end{align}
Here $U_{1,2}=\frac{U}{2}(1\pm\sqrt{1+4A/U^2})$ and $a$ is a real constant of integration.
In the following we set $x_0=0$ without loss of generality. 
Solving with respect to $u$ we have
\begin{align}
	u = \frac{U_1-U_2e^{i k(x-U t)+k a}}{1-e^{i k(x-U t)+k a}} \,,
\end{align}
where $k=-\frac{1}{2\nu_o}(U_1-U_2)$ and from analyticity of $u$ in the lower half plane we fix $a>0$. Fixing $k>0$, analytically continuing $x\to z=x+iy$ and requiring that $u\to 0 $ as $y\to -\infty$ we obtain $U_2=0$. This means that integration constant $A=0$ and $U_1=U$. 
We obtain the solution 
\begin{align}
	u=\frac{U}{1-e^{i k(x-U t)+k a}}	
\label{eq:uminus-app}
\end{align}
with $k>0$, $a>0$ and 
\begin{align}
	U=-2\nu_o k
\end{align} 
for $u$ consistent with boundary conditions and analyticity requirements.
The overall velocity field $\tilde{u}=u+\bar{u}$ is then given by,
\begin{align}
	\tilde{u}=-2\nu_o k\left(1-\frac{\sinh( k a)}{\cosh(k a)-\cos(k(x-Ut))}\right).
\end{align}
In the limit of $a\rightarrow +\infty$ while keeping $k>0$ fixed, we obtain,
\begin{align}
	\tilde{u}=-2\nu_o k(1-\tanh(ka))+2\nu_o k\frac{\tanh(k a)}{\cosh(ka)}\cos(k(x-Ut)).
\end{align}
The first term vanishes as $a\rightarrow +\infty$ for $k>0$. In this limit we recover the linearized Lamb's solution,
\begin{align}
	\tilde{u} \approx 4\nu_o ke^{-ka}\cos(k(x-Ut)).
\end{align}
We can now compute the non linear profile of the height function $h(x,t)$ using the kinematic condition of the form $h=\int^t \phi_y dt=-\int^t \tilde{u}^H dt$. We have defined 
$\tilde{u}^H=i(\bar{u}-u)$ and using the expression for $u$ in Eq.~\ref{eq:uminus} we get,
\begin{align}
	\tilde{u}^H=-2\nu_o k\frac{ \sin(k(x+2\nu_o k t))}{\cos(k(x+2\nu_o k t))-\cosh(ka)}
\end{align}
Using the above expression we can obtain $h(x,t)$ as,
\begin{align}
	h(x,t)=-\int^t \tilde{u}^H dt=\frac{1}{k}\log [\cosh (a k)-\cos (k (x+2\nu_o k t))]+\text{const.}	
\end{align}
The above expression was restricted to $k>0$. We can now recover the linearized Lamb limit by taking $a\rightarrow +\infty$. In this limit up to an overall constant we obtain,
\begin{align}
	h(x,t)\approx -2\frac{e^{-a k}}{k}\cos \Big(k (x+2\nu_o |k| t)\Big)\,.
\end{align}
The above expression is consistent with the linearized Lamb solutions as presented in the main text.

%%%%%%%%%%%%%%%%%%%%%%
\section{Multi-pole solutions for chiral Burgers equation}
 \la{app:multipole}
%%%%%%%%%%%%%%%%%%%%%%

In this section, we derive the multi-pole solution for the non-linear chiral Burgers equation,
\begin{align}
	u_t+2u u_x - 2i\nu_o u_{xx} = 0 \,,
 \la{eq:decoupvelapp}
\end{align}
where we additionally require that $u(x,t)$ is analytic in the lower half of complex plane after analytic continuation in $x$.
We make the following ansatz,
\begin{align}
	u(x,t)=-2i \nu_o\sum_{j=1}^N\frac{1}{x-z_j} 	
\end{align}
which is analytic in the lower half plane and all the poles $z_j$ are in the upper half plane. Substituting the above ansatz in the equation, we obtain a dynamical system for the poles,
\begin{align}
	\dot z_j=4i\nu_o \sum_{k\ne j}^N\frac{1}	{z_k-z_j}.
 \label{eq:multisolCB}
\end{align}
The above dynamical system corresponds to a valid solution of Eq.~\ref{eq:decoupvelapp} only for a finite time as at least one of the poles inevitably hits the real axis after which the the physical description through the chiral Burgers equations becomes untenable. In other words, beyond this time scale either higher order corrections become important or the boundary layer approximation breaks down. However, not all multipole solutions are unstable. In particular, we can arrange the poles such that it sums up to the non-linear periodic form given in Eq.~\ref{eq:uminus-app}. Consider $z_j(t)=2\pi j/k$ spaced evenly at distance $2\pi/k$. For
such an arrangement of poles, the multipole ansatz becomes,
\begin{align}
	u(x,t)=-\nu_o k- \frac{2i\nu_o}{x-U t-ia}-\sum_{n=1}^{\infty}\left(\frac{2i \nu_o}{(x-U t-ia)-\frac{2n\pi}{k}}+\frac{2i \nu_o}{(x-U t-ia)+\frac{2n\pi}{k}}\right).\end{align}
The constant first term is to satisfy the boundary condition $u\rightarrow 0$ as $z\rightarrow \infty$. The terms in the above expression sum to, 
\begin{align}
	u(x,t)=-\nu_o k \left(1+i \cot\left(\frac{k(x-Ut-ia)}{2}\right)\right).
	\end{align}
	The above expression can be recast in a more familiar form that reproduces equation \ref{eq:uminus-app},
\begin{align}
	u(x,t)= \frac{-2\nu_o k}{1-e^{i k(x-Ut)+k a}}.
\end{align}
The velocity $U=-2\nu_o k$ can be determined by regularizing the summation in Eq.~\ref{eq:multisolCB}. In next few sections we consider single and double pole solutions.  
%%%%%%%%%%%%%%%%%%%%%%
\subsection{Single-pole solution}

In a single pole solution $N=1$ the system (\ref{eq:multisolCB}) degenerates to $z_1=const$ and we have 
\begin{align}
	u(x,t)=-2i \nu_o\sum_{j=1}^N\frac{1}{x-ia} 	
\end{align}
to be a solution of (\ref{eq:decoupvelapp}). This solution is time-independent for $u(x,t)$ and corresponds to the real $\tilde{u}=u+\bar{u}$ given by
\begin{align}
	\tilde{u} = -\frac{4\nu_o a}{x^2+a^2}\,.
\end{align}
The profile of the wave $h(x,t)$ corresponding to this solution can be found as
\begin{align}
	h &= \frac{4\nu_o x}{x^2+a^2} t \,
\end{align}
and grows in time linearly. In time of the order of $t_* = a^2/\nu_o$ the curvature of the profile becomes significant and the assumptions of small curvature used in deriving (\ref{eq:almostBO-duplicate}) are violated.
%%%%%%%%%%%%%%%%%%%%%%
\subsection{Two-pole solution}

Let us consider an example of two poles. We have
\begin{align}
	\dot z_{1,2}(t) = \mp\frac{ 4i\nu_o}{z_1-z_2} \,.
 \label{eq:2soldyn1}
\end{align}
These equations are easy to solve to obtain
\begin{align}
	z_{1,2}(t) = z_0 \pm \sqrt{-4i\nu_o (t-t_0) +C} \,,
 \label{eq:2soldyn2}
\end{align}
where $\I(z_0)>0$ and $C$ is an arbitrary real constant. It is clear that at some time one of the poles approaches the real axis. At this point the assumptions of small nonlinearity used in the derivation of chiral Burgers equation are violated. A typical time scale for this is $t_*=(\I(z_0))^2/\nu_o$.
%%%%%%%
\begin{figure}
\centering
\includegraphics[scale=0.6]{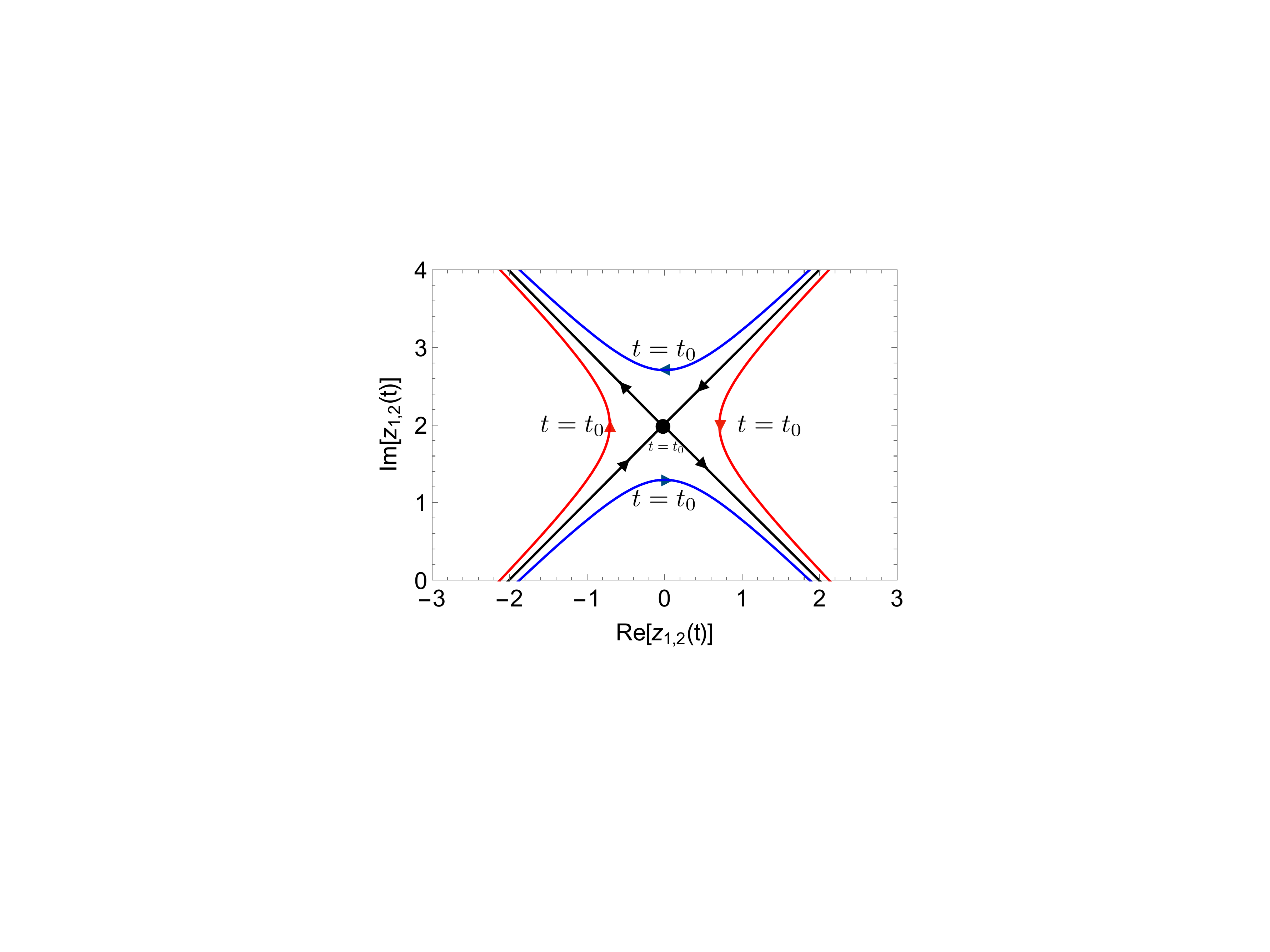}
\caption{\label{fig:2polepar} Dynamics of the two poles given by Eq.\ref{eq:2soldyn2} in the complex plane with time as the parameter. We have set $\nu_o=1$, $t_0=0$ and $z_0=i$. The three colors correspond to $C=-0.5$ (Blue), $C=0.5$(Red) and $C=0$ (Black).}
\end{figure}
%%%%%%%
%%%%%%%%%%%%%%%%%%%%%%
\section{General family of non-linear equations}
 \la{app:family}
%%%%%%%%%%%%%%%%%%%%%%

In this section, we put together a one parameter family of non-linear equations that contains both Benjamin-Davis-One (BDO) and chiral Burgers equation as a special cases. We introduce the following non-linear equation,
\begin{align}
u_t +2\lambda uu_x-2(1-\lambda)u^H u^H_x +2\nu_ou^H_{xx} = 0. 	
\label{eq:family}
\end{align}
In the above family of non-linear equations, we obtain BDO for $\lambda=1$ and chiral Burgers type equation for  $\lambda=1/2$. In the limiting case of $\lambda=0$, we obtain a new non-linear equation which possesses multi-soliton solutions,
\begin{align}
	u_t - 2u^H u^H_x +2\nu_ou^H_{xx} = 0.
\label{eq:boundbdo}
\end{align}
However, unlike the BDO case where multi-soliton solutions are scattering states, the multi-soliton solutions of Eq.~\ref{eq:boundbdo} form bound states. Generally, whether these solitons are of bound state type or scattering type is dictated by the parameter $\lambda$. Studying the multi-soliton dynamics of this general family sheds light on the dynamics of the chiral Burgers case which contains the physics of non-linear odd surface waves discussed in this work. One can write down a multi-pole solution for the equation \ref{eq:family} as,
\begin{align}
u(x,t)=-2i\nu_o \sum_{j=1}^N\frac{1}{x-z_j}+c.c.
\end{align}
Substituting the multi-pole solution in Eq.~\ref{eq:family}, we obtain the  dynamics of poles as,
\begin{align}
	\dot{z}_j = 4i\nu_o\left( \sum_{k=1, k\neq j}^N \frac{1}{z_k-z_j}
    +(1-2\lambda)\sum_{k=1}^N \frac{1}{\bar{z}_k-z_j}\right).
\end{align}
The above dynamical system corresponds to the celebrated integrable BDO equation for $\lambda=1$. The pole dynamics for the BDO corresponds to the scattering states of multi-soliton solutions. For $\lambda=0$ we have a new kind of multi-soliton solution corresponding to Eq.~\ref{eq:boundbdo} which form bound states of the multi-soliton solutions. The stability of these bound states is an interesting future direction. For $\lambda=1/2$, we recover chiral Burgers equation where the second term in the pole dynamics drops out and we obtain multi-pole dynamics of  Eq.~\ref{eq:multisolCB}.

\begin{comment}

\section{Incompressible, Dissipative, Parity-Violating Boundary waves}

\section{Incompressible and Dissipative odd waves}

Here we present the main equations for incompressible 2D fluid with odd viscosity \cite{avron1998odd,ganeshan2017odd}. We start with the equations of motion of an incompressible 2D fluid\footnote{We also assume that the density is constant, taken to be 1.}
\bea
	D_t v_i = \p_j T_{ij}\,,
 \la{eq:euler1}
\eea
where $D_t=\p_t +\mathbf{v}\cdot\bm\nabla$ is a material time derivative and the stress tensor is given by
\bea
	T_{ij} &=& -p\delta_{ij} +\nu_e(\p_i v_j+\p_j v_i) +\nu_o (\p_i^*v_j+\p_i v_j^*)\,.
 \la{eq:stress1}
\eea
Here we used the convenient notation $a_i^*=\epsilon_{ij}a_j$ and the last term is the odd viscosity contribution to the stress. For the incompressible fluid $\bm\nabla\cdot \mathbf{v} =0$.

We can rewrite (\ref{eq:euler1}) with (\ref{eq:stress1}) as
\bea
	D_t \mathbf{v} = -\bm\nabla \tilde{p} +\nu_e \Delta \mathbf{v}\,,
 \la{eq:euler2}
\eea
where the modified pressure $\tilde{p}$ is expressed through the vorticity $\omega=\epsilon_{ij}\p_iv_j$ as $\tilde p = p -\nu_o \omega$. We notice that the pressure of an incompressible fluid is not a state variable but is determined by the fluid flow. Therefore, the equation (\ref{eq:euler2}) along with the incompressibility condition essentially do not depend on the odd viscosity $\nu_o$ \cite{avron1998odd,ganeshan2017odd}. Indeed, let us take the curl of (\ref{eq:euler2}). We obtain the conventional equation for the vorticity of the incompressible 2D fluid
\bea
	\p_t\omega +\bm\nabla(\omega\mathbf{v}) = \nu_e\Delta\omega\,.
 \la{eq:omega1}
\eea
The equation (\ref{eq:omega1}) does not contain $\nu_o$ and are sufficient to determine both components of velocity field. The pressure of the fluid can be found afterwards using (\ref{eq:euler2}) and the known velocity field. 

The dependence on odd viscosity, however, can occur through boundary conditions \cite{ganeshan2017odd}. We will be interested in free surface boundary conditions.

In this section we generalize Lamb's solution \cite{lamb1932hydrodynamics} for long gravity waves in a fluid with shear viscosity to the case where odd viscosity is present. The free surface (no-stress) dynamic boundary conditions (DBC), as well as kinematic boundary conditions (KBC) can be written as 
\begin{align}
	T_{ij} n_{j} = 0\,	, \quad \partial_{t} h(x,t) + v_{x} \partial_{x} h(x,t) = v_{y}\,
\end{align}
where we parameterized the surface by the function $h(x,t)$, with normal vector $\mathbf{n} = (-h_{x}, 1) / \sqrt{1 + (h_x)^{2}}$. Assuming small amplitude, long wavelength waves and linearizing these boundary conditions and using (\ref{eq:f1}) we obtain at the surface 
\bea
	\tilde{p} &=& 2\nu_e \p_y v_y -2\nu_o \p_x v_y\,,
 \la{eq:dbcy} \\
 	0 &=& \nu_e (\p_x v_y +\p_y v_x) +2\nu_o \p_y v_y \,,
 \la{eq:dbcx}\\
 	h_t &=& v_y\,.
 \la{eq:kbc}
\eea
The bulk equations for $y\leq h(x,t)$ after linearization are (we also add gravitational potential $gy$):
\bea
	\bm\nabla\cdot\mathbf{v} &=& 0\,,
 \\
 	\p_t \mathbf{v} &=& -\bm\nabla (\tilde{p}+gy) +\nu_e\Delta\mathbf{v}\,,
 \\
 	\p_t \omega &=& \nu_e \Delta \omega\,.
\eea   
The last equation is the consequence of the first two. 

\subsection{No solutions exist for $\nu_{e} = 0$}

We may refer to this setting as odd Euler flow. It turns out that linear propagating waves are not supported in this setting. This can be seen ....

The main less is that irrotational flow set in motion will develop a transverse stress that accelerates the boundary, resulting in a vorticity layer.

\subsection{Weak solutions and the $\nu_{e} \to 0$ limit}

In this section, we study the dissipationless limit $\nu_{e} \to 0$. This is a sensible limit to take in considering the consistency equation for the dispersion relation. We obtain from it to leading order 

\begin{align}
\Omega^{2}+ 2 a^{2} k |k| \Omega - g |k|  = 0
\end{align}

However, the flow does not have a similarly smooth limit. Indeed, the penetration depth of the vorticity layer tends to zero as $\nu_{e} \to 0$, which implies that the absolute magnitude of the vorticity blows up.

\section{Non-linear Boundary waves}

We begin with a linear analysis which closely follows Lamb's solution of damped gravity waves \cite{lamb1932hydrodynamics}. Specifically, we consider a semi-infinite fluid occupying the lower half plane ($y <0$), subject to the vertical ``gravitational" force, which we denote by $f_{y}$.  

The equations of motion for a 2D incompressible fluid with even and odd viscosity is encapsulated in the complex notation

\begin{align}
D_{t} v_{x} &= - \partial_{x} p + \nu_{e} \Delta v_{x} + \nu_{o} \Delta v_{y} + f_{x}\\
D_{t} v_{y} &= - \partial_{y}p + \nu_{e} \Delta v_{y} - \nu_{o} \Delta v_{x} - f_{y}\\
\partial_{x} v_{x} &+ \partial_{y}v_{y} = 0
\end{align}
The stress-tensor is
\begin{align}
T_{ij} = - p \delta_{ij}  + \nu_{e} (\partial_{i} v_{j} + \partial_{j} v_{i}) + \nu_{o} (\partial_{i}^{*} v_{j} + \partial_{i} v_{j}^{*})
\end{align}

For the problem at hand, we require stress-free boundary conditions, as well as kinematic boundary conditions (KBC). Parameterizing the surface by the function $h(x,t)$, with normal vector $n = (-h_{x}, 1) / \sqrt{1 + (h')^{2}}$, the kinematic and dynamic boundary conditions (DBC) then read

\begin{align}
T_{ij} n_{j} = 0	, \quad \partial_{t} h(x,t) + v_{x} \partial_{x} h(x,t) = v_{y}
\end{align}

Incompressibility allows one to solve for the flow directly. The equation for vorticity transport implies that the vorticity will be independent of odd viscosity. However, the pressure will be shifted by an amount $\nu_{o}\omega$. The linearized equations will then have the following solutions which decay in the interior of the fluid $y \to - \infty$

\begin{align}
v_{x} &= e^{ i k x - i \Omega t} \left( A e^{ |k| y} + B e^{ my}\right)\\
v_{y} &= e^{ i k x - i \Omega t} \left( -i\frac{k}{|k|} A e^{ |k| y} -i \frac{k}{m} Be^{ m y}\right)\\
p & = -f_{y}y + 	e^{ i k x - i \Omega t}\left( \frac{\Omega}{k} A e^{ |k| y} + \frac{i\nu_{o} \Omega}{\nu_{e} m} B e^{ m y}\right)\\
p & = -f_{y}y + 	e^{ i k x - i \Omega t} \frac{\Omega}{k} A e^{ |k| y} + \nu_{o} \omega \\
m^{2} &= k^{2} - \frac{i \Omega}{\nu_{e}}
\end{align}

for reference, we have the vorticity

\begin{align}
\omega  = \frac{i \Omega}{\nu_{e} m} B e^{ i k x - i \Omega t + my}	
\end{align}

The coefficients $A$ and $B$ will be fixed by the boundary conditions. In components, and keeping only terms of order $1$ in $h$, the dynamical boundary conditions simplify to

\begin{align}
T_{yy} &= - p + 2\nu_{e}\partial_{y} v_{y} - \nu_{o}(\partial_{x} v_{y} + \partial_{y} v_{x}) = 0 \label{Tyy}\\
T_{xy} &= \nu_{e} (\partial_{x} v_{y} + \partial_{y} v_{x}) + 2\nu_{o}\partial_{y} v_{y} =  0	 \label{Txy}
\end{align}

Next, we take the time derivative of (\ref{Tyy}) and utilize the KBC to leading order $dy/dt = v_{y}$ that appears in the pressure. The result, defining for shorthand $p + f_{y}y = P$, is

\begin{align}
\frac{if_{y} 	}{ \omega}v_{y}   - P + 2 \nu_{e} \partial_{y} v_{y} - \nu_{o} (\partial_{x} v_{y} + \partial_{y} v_{x}) = 0
\end{align}

The DBC can be combined in a convenient form $T_{yy}+ i T_{xy} = 0$, using the complex viscosity $\nu = \nu_{e} + i \nu_{o}$. After accounting for the KBC, we get the holomorphic and anti-holomorphic components to read

\begin{align}
\frac{i g}{\omega} v_{y} - P + \nu \left( 2 \partial_{y} v_{y} + i \partial_{x} v_{y} + i \partial_{y} v_{x}\right) = 0	\\
\frac{i g}{\omega} v_{y} - P + \bar{\nu} \left( 2 \partial_{y} v_{y} - i \partial_{x} v_{y} - i \partial_{y} v_{x}\right) = 0	
\end{align}

For real $\nu$, this reduces to the system of equations satisfied in the absence of odd viscosity $\nu_{o}$. Carrying out the derivatives and evaluating this expression at $y = 0$ yields the following condition 

%\begin{align}
%\frac{i g}{\omega} \left( - i \frac{k}{|k|} A - \frac{i k }{m}B\right) - \frac{\omega}{k} A - \frac{i \nu_{o} \omega}{\nu_{e} m} B + \nu \Big(  2 i (|k| - k) A + i \frac{(m-k)^{2}}{m} B \Big) = 0\\
%\frac{i g}{\omega} \left( - i \frac{k}{|k|} A - \frac{i k }{m}B\right) - \frac{\omega}{k} A - \frac{i \nu_{o} \omega}{\nu_{e} m} B + \bar{\nu} \left( - 2 i (k + |k|) A - i \frac{(m+k)^{2}}{m} B \right)
%\end{align}

%\begin{align}
%\left[ \frac{gk}{\omega |k|} - \frac{\omega}{k} + 2 i \nu ( |k| - k) 	\right] \left[ \frac{g k}{m \omega} - i \bar{\nu} \frac{(m+k)^{2}}{m} - \frac{ i \omega \nu_{o}}{m \nu_{e}} \right] = \left[ \frac{gk}{\omega |k|} - \frac{\omega}{k} -2 i \bar{\nu} ( |k| + k) 	\right] \left[ \frac{g k}{m \omega} + i \nu \frac{(m-k)^{2}}{m} - \frac{ i \omega \nu_{o}}{m \nu_{e}} \right]
%\end{align}

%\begin{align}
% \Big[ g|k| - \omega^{2} + 2 i \nu ( |k| - k) k \omega 	\Big] \left[ gk - i \bar{\nu} (m+k)^{2} \omega - \frac{ i \nu_{o}}{ \nu_{e}}\omega^{2}  \right] = \left[ g|k| - \omega^{2} -2 i \bar{\nu} ( |k| + k)k \omega 	\right] \left[ g k + i \nu (m-k)^{2} \omega - \frac{ i \nu_{o}}{\nu_{e}} \omega^{2} \right]
%\end{align}

\begin{align}
\frac{ \Big( g|k| - \omega^{2} + 2 i \nu ( |k| - k) k \omega 	\Big) \Big( gk - i \bar{\nu} (m+k)^{2} \omega - i (\nu_{o}/\nu_{e})\omega^{2}  \Big) }{\Big( g|k| - \omega^{2} -2 i \bar{\nu} ( |k| + k)k \omega 	\Big)\Big( g k + i \nu (m-k)^{2} \omega -  i (\nu_{o}/\nu_{e}) \omega^{2} \Big)} = 1
\end{align}

An alternative form is

\begin{align}
\left( 1 - \frac{i \omega}{2k^{2} \nu_{e}} + \frac{i m \nu_{o}}{k \nu_{e}}\right) \left( 1 - \frac{i \omega}{2 k^{2} \nu}\right) + \frac{g|k|}{2 k^{2} \nu_{e}} 	\left( \frac{1}{2k^{2} \nu} + \frac{\nu_{o}}{\omega \nu} - \frac{\nu_{o} m}{\omega \nu k}\right) = \frac{m}{|k|} + i \frac{\nu_{o} k}{\nu_{e}|k|}
\end{align}

In dimensionless units, we introduce a horizontal $x$- scale $L$, a characteristic velocity $u_{0}$, and rewrite the equation in terms of dimensionless variables
\begin{align}
q &= |k| L, \quad \omega' = \frac{L}{u_{0}} \omega , \quad a = \frac{1}{A} = \frac{\nu_{o}}{u_{0} L}, \quad r = \frac{1}{R} = \frac{\nu_{e}}{u_{0} L}, \quad \nu' = r + i a, \quad \bar{\nu}' = r - i a \\
v_{g} &= \sqrt{g L}, \quad g' = \frac{v_{g}}{u_{0}} ,\quad m' = L m = L |k| \sqrt{1 - \frac{i \omega}{\nu_{e}k^{2}} } = q \sqrt{1 - \frac{i  \omega' }{r q^{2}}}\\
\chi &= \frac{k}{|k|}
\end{align}

Now dropping the primed notation we have in dimensionless quantities defined above

\begin{align}
&\frac{ \Big(  g q  - \omega^{2} + 2 i \nu  ( \chi  - 1) q^{2} \omega 	\Big) \Big(  g \chi q-   i \bar{\nu} (m+\chi q)^{2} \omega - i (a/r) \omega^{2}  \Big) }{\Big( g q-  \omega^{2}  -2 i \bar{\nu} ( \chi + 1)q^{2} \omega 	\Big)\Big(  g \chi q+  i \nu  (m-\chi q)^{2} \omega - i (a/r) \omega^{2} \Big)} = 1\\
\end{align}

\begin{align}
\left( 1 - \frac{i \Omega}{2\eta_{e}} + \frac{i \alpha a }{\eta_{e}}\right) \left( 1 - \frac{i \Omega}{2  \eta}\right) + \frac{1}{2 \eta_{e}} 	\left( \frac{1}{2 \eta } + \frac{\eta_{o}(1- \alpha)}{\Omega \eta} \right) = \frac{k}{|k|} \left( \alpha  + i \frac{\eta_{o} }{\eta_{e}}\right)
\end{align}

For $f_{y}= 0$, this becomes

\begin{align}
\frac{ \Big(  \omega - 2 i \nu ( |k| - k) k  	\Big) \Big(  i \bar{\nu} (m+k)^{2}  + i (\nu_{o}/\nu_{e})\omega  \Big) }{\Big(   \omega + 2 i \bar{\nu} ( |k| + k)k  	\Big)\Big(  - i \nu (m-k)^{2}  +  i (\nu_{o}/\nu_{e}) \omega \Big)} = 1
\end{align}

A convenient polynomial expression particular well suited for asymptotic analysis is given by

\begin{align}
%&\left( \omega^{3} + \left( 4 i r + 2 a \chi\right)q^{2} \omega^{2} + \left( - g q - 4 (r^{2} + a^{2}) q^{4}\right) \omega + 2 a \chi g q^{3} \right)^{2} = 4 q^{2} \chi^{2} \left( q^{2} - \frac{i \omega}{r}\right) \left( a g q - a \omega^{2} - 2\chi |\nu|^{2} q^{2} \omega \right)	^{2}\\
\left( \omega^{3} + \left( 4 i r + 2 a \chi\right)q^{2} \omega^{2} - \left(  g q + 4 |\nu|^{2} q^{4}\right) \omega + 2 a \chi g q^{3} \right)^{2} = - \frac{i}{r}4 q^{2} \left(\omega +i r q^{2} \right) \left(  a \omega^{2} -a g q  +2 \chi |\nu|^{2} q^{2} \omega \right)	^{2}
\end{align}

When $f_{y} = 0$, the consistency equation simplifies remarkably. Defining $v = \Omega / |k|^{2}$, we get

\begin{align}
v^{3} + 4 i \bar{\nu} \left( 1 + \frac{\nu}{r}\right) v^{2} + \left( 4 \tilde{a}^{2} - 8 \bar{\nu}^{2} - \frac{16 |\nu|^{2} \bar{\nu}}{r}\right) v + 16 i |\nu|^{2} \bar{\nu} \left( \frac{\nu}{r} - 2 \right) = 0	
\end{align}

Which implies the dispersion is quadratic with only the coefficient being affected by the presence of odd viscosity.

\subsection{Incompressible bulk modes}

\begin{align}
\dot{v}_{i} = - \nabla_{i} p + \nu_{o} \Delta v_{i}^{*}\\
\partial_{i} v_{i} = 0	
\end{align}

Implies

\begin{align}
\dot{\omega} = 0\\
\Delta p = \nu_{o} \Delta \omega 
\end{align}

The equations of motion are just

\begin{align}
\dot{v}_{i} = - \partial_{i} \left( p - \nu_{o} \omega \right)	
\end{align}

So that $\ddot{{\bf v}} = 0$, which means there are no propagating modes. This is simply a consequence of incompressibility. However, what we do get is apparently some sort of diffusion

\begin{align}
r_{i} = - \left(\partial_{i} \tilde{p} \right) t	
\end{align}

If the flow starts initially with some vorticity which is spatially

%%%%%%%%%%%%%%%%%%%%%%%%%%%
%%%%%%%%%%%%%%%%%%%%%%%%%%%%
\subsection{Validity of linearized solutions}

We are now in a position to quantify different length scales in the problem. Note that these solutions are obtained by neglecting the advective acceleration term $(\bf v. \bf{\nabla}) v$ in the equation of motion in Eq.~\ref{eq:lineareommain}. Consequently these are not exact solutions, but the required accuracy can be obtained by decreasing the amplitude of the wave. Consider the typical amplitude of the wave to be $a\propto (A, B)$. The dimensionless small parameter that provides the necessary accuracy in the linearized limit is given by $a/\lambda$, where $\lambda=2\pi/k$ is the wavelength.

In the presence of the $\nu_e$, there exists another length scale coming from the exponential decay term $e^{my}$. This length scale is given by $\delta=1/m$. The vorticity of the fluid given in Eq.~\ref{eq:linvort} decays from the boundary to the bulk dictated by the inverse length scale $m^2=k^2-i\Omega/\nu_e$. For the $\Omega=0$ branch this scale is again set by the wavenumber $m=\pm |k|$. However, for the $\Omega =-2\nu_o k|k|$ branch we obtain an interesting scaling behavior in the inviscid limit $\nu_e\rightarrow 0$ where $m\sim-e^{-i\pi/4}\sqrt{\nu_o/\nu_e}k$. This length scale determines the thickness of a {\it boundary layer} where most of the vorticity of the fluid is confined. Outside this layer we can approximately neglect the vorticity and can treat the fluid as irrotational and use the potential flow analysis done in Sec.~\ref{sec:potflow}. 

A natural question then arises, if the dropping the advective acceleration term $(\bf v. \bf{\nabla}) v$ from the equations of motion is justified in the presence of boundary layer. If yes then what is the regime of validity of the linearized solutions in two scales of interest $\delta<a<\lambda$ and $a<\delta<\lambda$.

%%%%%%%%%%%%%%%%%%%%%%%%%%%
%%%%%%%%%%%%%%%%%%%%%%%%%%%%
\subsection{Inviscid limit of linear odd surface waves in the absence of gravity}

{\blue SASHA: this section and the table should go to appendix or removed altogether. }

 We now take the inviscid limit for the modified Lamb's solutions and check if these solutions satisfy the boundary conditions written in the inviscid limit. The two DBC in the $\nu_e \rightarrow 0$ limit reduces to,
\begin{align}
 \tilde p=- 2\nu_{o}\partial_{x} v_{y}|_{y=0},\quad \p_x v_x=0|_{y=0}.
\end{align}
In this limit, the consistency equations in Eqs.~(\ref{eq:AB1} and \ref{eq:AB2}) simplify to $B=0$ and $A|k|+B m=0$ respectively.   $A|k|+B m=0$ implies $v_x=0$ at the boundary $y=0$. This is indeed the tangential stress conditions at the boundary $\p_x v_x=0|_{y=0}$. Notice that $\omega\ne 0$ ($B\ne 0$) is absolutely necessary to satisfy this boundary condition in a non-trivial way. The second step is to satisfy the normal DBC $\tilde p=- 2\nu_{o}\partial_{x} v_{y}|_{y=0}$ which is easily satisfied by directly substituting $B=0$ in Eqs.~\ref{eq:linvx} and \ref{eq:pressure}. The normal stress DBC in the inviscid limit behaves as if the fluid is irrotational and thus the dispersion relation obtained is unaffected by the presence of boundary vorticity that was absolutely essential to satisfy the tangential DBC. Thus the modified Lamb's solutions in the inviscid limit can be reinterpreted as irrotational motion in the bulk with a thin boundary layer at the boundary whose sole purpose in the linear hydro limit is to satisfy the tangential stress boundary condition. The thickness of this boundary layer in the inviscid limit given by,
\begin{align}
	\delta\sim m^{-1}\sim \sqrt{\frac{\nu_e}{\nu_o}}\lambda.
\end{align}
The vorticity that is confined at the boundary is necessary for consistency of the flow solutions to satisfy the boundary conditions. The presence of an infinitesimal layer of vorticity is required for the existence of dispersing odd surface waves given in Eq.~\ref{eq:dispmain} even though the dispersion equations itself remains unmodified by this boundary layer.  Recall that the well known dispersion relation for the gravity waves are given by $\Omega_{gw}=\sqrt{gk}$, where $g$ is the acceleration due o gravity. The thickness of the  boundary layer for this case is given by $\delta_{gw}\sim \lambda^{1/4}(\sqrt{\nu_e/\sqrt{g}})$. The gravity acts as the dominant irrotational part in the bulk and the boundary layer vorticity is again determined by the infinitesimal $\nu_e$. The existence of linearized solutions in different limiting case can be summarized in Table.~\ref{tab:incom}.

\begin{table}
\begin{center}
	\begin{tabular}{|c|c|c|c|p{7cm}|}
	\hline
 g & $\nu_e$ & $\nu_o$ & $\Omega (k)$ & Incompressible linear flow type\\
  \hline
  $\ne 0$ & =0 & =0 & $\sqrt{gk}$ & Fully irrotational \\
    \hline
    $\ne 0$ & $\rightarrow 0$ & = 0 & $\sqrt{gk}\pm 2i\nu_e k^2$ & Irrotational bulk with boundary layer $\delta_{gw}\sim \lambda^{1/4}(\sqrt{\nu_e/\sqrt{g}})$ \\
    \hline
  =0 & $\rightarrow 0$ & $\ne 0$ & $0, -2\nu_o k|k|$ & Irrotational bulk with boundary layer $\delta\sim \sqrt{\frac{\nu_e}{\nu_o}}\lambda$\\
  \hline
   =0 or $\ne 0$ & $=0$ & $\ne 0$ & No Solutions & No Solutions\\
   \hline
     =0 & $\rightarrow 0, \ne 0$ & $=0$ & No Solutions & No Solutions\\
  \hline
\end{tabular}
\end{center}
\caption{This table summarizes the solutions and surface wave propagation of a two dimensional incompressible fluids with three parameters, namely, acceleration due to gravity $g$, dissipative viscosity $\nu_e$ and odd dissipation less odd viscosity $\nu_o$. The table summarizes linearized hydrodynamic solutions and the dispersion relation of the surface wave propagation in different limiting cases.}
\label{tab:incom}
\end{table}

%%%%%%%%%%%%%%%%%%%%%%%%%%%%
\subsection{Free surface boundary conditions }

The free surface (no-stress) dynamic boundary conditions (DBC), as well as kinematic boundary conditions (KBC) can be written as 
\begin{align}
	T_{ij} n_{j} = 0\,	, \quad \partial_{t} h(x,t) + v_{x} \partial_{x} h(x,t) = v_{y}\,
\end{align}
where we parameterized the surface by the function $h(x,t)$, with normal vector $\mathbf{n} = (-h_{x}, 1) / \sqrt{1 + (h_x)^{2}}$. It will also be convenient to introduce the unit vector tangent to the surface in directed in the counterclockwise direction $\mathbf{s}=-\mathbf{n}^*=-(1,h_{x}) / \sqrt{1 + (h_x)^{2}}$.

It is useful for the following to rewrite KBC and DBC in the orthogonal coordinate systems attached to the surface of the fluid. Namely, we will introduce coordinates $(s,n)$ in which unit vectors are given by $\mathbf{n}=(0,1)$ and $\mathbf{s}=(1,0)$ at the surface of the fluid which can be parameterized by $(s,0)$. We have at the boundary
\begin{align}
	\p_n \mathbf{n}=\p_n\mathbf{s} = 0\,,\quad \p_s \mathbf{s} = -\kappa \mathbf{n}\,, \quad\p_s \mathbf{n} = \kappa\mathbf{s}\,.
\end{align}
Here we introduce the curvature of the boundary $\kappa = h_{xx}(1+h_x^2)^{-3/2}$. KBC in new coordinates can be written as $\mathbf{v}=(x_t(s,t),h_t(s,t))$ or
\bea
	\p_t\mathbf{s}(s,t) = -\mathbf{n}(\p_sv_n-\kappa v_s) \,.
\eea

Let us now write down DBC in new coordinates. We have
\bea
	T_{nn} &=& -p +2\nu_e \p_n v_n -\nu_o (\p_s v_n+\p_nv_s-\kappa v_s) = 0\,,
 \nonumber \\
 	T_{sn} &=& \nu_e(\p_s v_n+\p_nv_s-\kappa v_s) +\nu_o (\p_s v_s-\p_nv_n+\kappa v_n)= 0 
 \nonumber
\eea
In new coordinates close to the boundary the condition for being incompressible and irrotational become 
\begin{align}
	C=\mathbf{\nabla}\cdot\mathbf{v} = \p_s v_s+\p_n v_n+ \kappa v_n = 0\,,\quad
 	\omega=\mathbf{\nabla}\times\mathbf{v} = \p_n v_s-\p_s v_n+ \kappa v_s = 0\,,
\end{align}
and stress components can be simplified to rewrite the boundary conditions as
\begin{align}
	T_{nn} = -p +2\nu_e \p_n v_n +2\nu_o \p_nv_s=0 \,,\quad
 	T_{sn} = 2\nu_e\p_n v_s -2\nu_o \p_n v_n=0 \,,\quad \p_t\mathbf{s}(s,t) = -\mathbf{n}\p_nv_s
\end{align}

%%%%%%%%%%%%%%%%%%
%%%%%%%%%%%%%%%%%%
%%%%%%%%%%%%%%%%%%
%%%%%%%%%%%%%%%%%%
%%%%%%%%%%%%%%%%%%

Let us consider a particular example of the incompressible and irrotational flow in the domain $y\leq 0$. Then, from (\ref{eq:fi}) assuming that $\mathbf{n}=\hat{\mathbf{y}}$ is a unit vector in positive $y$ direction, we obtain 
\bea
	f_x/\rho_0 &=& \rho_0 (-2\nu_e\phi_{xx}^H-2\nu_o\phi_{xx}) \,,
 \nonumber \\
 	f_y/\rho_0 &=& -p/\rho_0 -2\nu_e\phi_{xx} +2\nu_o \phi_{xx}^H\,,
 \nonumber
\eea
where we used $\phi_x=\phi_y^H$ and $\phi_y=-\phi_x^H$ and the Hilbert transform is defined as
\bea
	\phi^H(x) = \frac{1}{\pi}P.V.\int \frac{\phi(x')\,dx'}{x'-x}.
\eea

%%%%%%%%%%

Let us consider the vorticity free fluid on a domain $B$ homeomorphic to the half-plane. We are interested in the motion of the boundary of this domain. We assume that vorticity is zero in the bulk and introduce the velocity potential $\bm{v}=\bm{\nabla}\phi$. The incompressibility condition becomes
\bea
	\Delta \phi = 0\,.
\eea
We solve equations of motion in the bulk
\bea
	\tilde{p} +\phi_t +\frac{1}{2}(\bm\nabla \phi)^2 = const
\eea
and impose the boundary condition (\ref{eq:modp}) on the boundary of the domain $\Gamma =\p B$
\bea
	\phi_t +\frac{1}{2}(\bm\nabla \phi)^2
	+2\nu_o\Big(\p_s v_n +\kappa \int^s ds\, \kappa v_n \Big)
	=0\,.
\eea
This equation together with the condition of harmonicity of $\phi$ and kinematic boundary condition determines the evolution of the boundary in time. 

It is convenient to parametrize the boundary in terms of the conformal mapping $z=w(\zeta,t)$ from the upper half-plane to the domain $B$. We assume $w(\zeta)\to \zeta $ at $\zeta\to \infty$. We would like to formulate all equations in $\zeta$-plane. We have
\bea
	\p_s &=&  |w'|^{-1}(\p_\zeta+\bar{\p}_\zeta)\,,
 \\
 	\p_n &=&  i |w'|^{-1}(\p_\zeta-\bar{\p}_\zeta)\,.
\eea
The velocity components correspondingly
\bea
	v_s &=&  |w'|^{-1}(\p_\zeta \phi+\bar{\p}_\zeta \phi)\,,
 \\
 	v_n &=&  i |w'|^{-1}(\p_\zeta\phi-\bar{\p}_\zeta\phi)\,.
\eea
For time derivative
\bea
	\phi_t\Big|_z = \phi_t\Big|_\zeta +\frac{\dot{w}}{w'} \p_\zeta\phi +\frac{\dot{\bar{w}}}{\bar{w}'}\bar{\p}_\zeta \phi\,.
\eea
The curvature
\bea
	\kappa =- \frac{i}{2|w'|}\left[\frac{w''}{w'}-\frac{\bar{w}''}{\bar{w}'}\right]
\eea

The kinematic boundary condition relates $w(\zeta,t)$ and $\phi(\zeta,t)$ so that at the boundary $\zeta_2=0$ we have
\bea
	\R[\overline{i\p_1 w} \, \p_t w(\zeta_1,t)] = \R[\overline{i\p_1 w} \, (-2i \p_{\bar{z}}\phi]
\eea
or
\bea
	\p_1\phi = -\I [\p_1\bar{w}\,\p_t w] \,.
\eea
Taking
\bea
	w(\zeta,t) = \zeta + f(\zeta,t)
\eea
we have
\bea
	\p_1\phi = -\I [\p_t f(1+\p_1 \bar{f})] \,
\eea
and splitting $\phi$ into analytic and antianalitic parts
\bea
	\phi =\phi^+(\zeta,t) +\phi^-(\bar\zeta,t)
\eea
we obtain
\bea
	\p_1\phi^+ = -\frac{i}{2}\p_t f (1+\p_1f)\,.
\eea

%\bea
%	w' &=& A (\phi_x -i\phi_x^H)+1 \,,
% \\
% 	\phi &=& \frac{1}{2A}(w-\zeta) +c.c.
%\eea
%where $\phi_x=\p\phi/\p\zeta_1$. Then
%\bea
%	\kappa = \frac{\phi_{xx}\phi_x^H-\phi_{xx}^H\phi_x}{(\phi_x^2+{\phi_x^H}^2)^{3/2}}
%\eea
%and

The dynamic boundary condition becomes in $\zeta$-plane
\bea
	\phi_t +\frac{\p_t w}{w'}\p_1\phi^++\frac{\p_t \bar w}{\bar w'}\p_1\phi^-
	+\frac{2}{|w'|^2}|\p_1\phi^+|^2
	+2\nu_o\left[\p_s v_n +\kappa \int^s ds\, \kappa v_n \right] =0\,,
\eea
where 
\bea
	v_n &=& \frac{1}{|w'|}\p_1\phi^H\,,
 \\
	\p_s v_n &=& \frac{1}{|w'|}\p_1 \frac{\phi_1^H}{|w'|} \,.
\eea
In linear approximation we have
\bea
	\phi_t +2\nu_o\phi_{11}^H = 0 \,.
\eea
Keeping quadratic terms we have
\bea
	\phi_t -2\phi_1^H\phi_1+ \frac{1}{2}(\phi_1^2+{\phi_1^H}^2) +2\nu_o\phi_{11}^H = 0\,.
\eea

%\begin{widetext} 
%This is where widetext begins
%\end{widetext} 

%%%%%%%%%%%%%%%%%%%%%%%%%%%%
%%%%%%%%%%%%%%%%%%%%%%%%%%%%
\section{Main results }
%%%%%%%%%%%%%%%%%%%%%%%%%%%%

{\bf Dispersion and boundary layers:} Within the linearized hydrodynamics the dispersion of the surface waves is completely determined by the odd viscosity term and is given by,
\begin{align}
\Omega=-2\nu_o k |k|.
\label{eq:disp}	
\end{align}
Where $\Omega$ is the frequency and $k$ is the wave number associated with the linear oscillating wave. Notice that the above dispersion relation corresponds to a unidirectional (chiral) propagating mode which indicates parity violation. {\blue NOTE: Need to add all the relevant scales comparing $\nu_o\gg (\nu_e, 1/v_s)$. Compare to the known case of gravity waves with dissipation. Add $\Omega=0$ branch discussion}. All the other relevant parameters associated to dissipation ($\nu_e$) and  compressibility ($v_s$) enters the dispersion at the higher order. We further show that the physical implications of the odd viscosity is confined to the dispersion of these surface waves which we dub as {\it odd surface waves}. We show that the {\it odd surface waves} propagate in the background of a bulk potential (irrotational) flow with a thin boundary layer at the surface where most of the vorticity is confined. The only role of this boundary layer is to satisfy the tangential stress conditions given in Eq.~\ref{eq:lineardbcT}. The thickness $\delta$ of this boundary layer scales as $\delta\sim \sqrt{\nu_e/\nu_o}$.

{\bf Non-linear effects:} The interplay of dissipation and compressibility is confined to the boundary layer as the bulk solutions are completely determined by the potential flow. The physical structure of the boundary layer only manifests in the non-linear terms. We show that the tangential stress can be satisfied by two completely different type of boundary layers. The extremities manifest in the parametric regimes $\nu_e=0, v_s\rightarrow \infty$ and $\nu_e\rightarrow 0, v_s=\infty$, where the former corresponds to the case of compressible boundary layer attached to an incompressible bulk potential flow. The later limit corresponds to a dissipative incompressible boundary layer attached to a dissipation-less incompressible bulk. These two completely different mechanisms supply the necessary vorticity that satisfies the tangential stress conditions. Remarkably, the normal stress condition is oblivious to these mechanisms at the linear order rendering the same dispersion  Eq.~\ref{eq:disp} in the two limits. We also derive the general solutions and wave dispersion that interpolates between these two regimes.

{\blue SASHA: The following piece (until the end of the section should be either removed or used somewhere else. It is not needed to proceed to linear solution.}

In the absence of viscosities the $T_{sn}=0$ boundary condition is trivially satisfied and the normal stress boundary condition $T_{nn}=p=0$ together with KBC determine the dispersion of the surface waves. However, if any of the viscosity coefficients is non-vanishing it is impossible to satisfy both dynamical boundary conditions for incompressible and irrotational fluid even when viscosity is very small. Notice that the equation of motions do not see this inconsistency since the $C=0, \omega=0$ solution negates the effects of viscous acceleration in the bulk of the fluid.  However, at the boundary, even though the viscous forces are zero, the viscous stress terms are non-zero for a  $C=0, \omega=0$ solution. The DBC requires the viscous stress terms to vanish at the boundary. This mismatch in the viscous stress terms near the boundary results in a fluid acceleration confined to a layer of the size of $\delta\sim\sqrt{\nu_e}$. This thin layer is the well known {\emph boundary layer} where all the fluid vorticity is confined which is consistent with the vanishing tangential stress term consistent with the DBC. Outside this boundary layer the fluid is completely defined by a potential flow solution.  Thus the surface wave problem for the fluid with an infinitesimal viscosity can be understood as a irrotational and incompressible bulk fluid with a thin boundary layer that carries the vorticity. This boundary layer behaves like a rigid object that is stuck at the surface of the fluid and moves with surface motion. 

In the following sections, we demonstrate how the surface dynamics in the presence of the boundary layers is influenced by the presence of  the odd viscosity $\nu_o$. In particular, we show that the thickness of the boundary layer scaling is dramatically influenced by the presence of odd viscosity. We also show that there exists two distinct mechanisms in the form of dissipation and compressibility that can supply the necessary vorticity at the boundary layer. We begin by deriving Lamb type linearized solutions for the incompressible surface waves for the case $\nu_e\ne 0, \nu_0\ne 0, C=0$.

\end{document}